\documentclass[nofootinbib,superscriptaddress]{revtex4}

\usepackage[percent]{overpic}

\usepackage{natbib}
\usepackage{graphicx}
\usepackage{epsfig}
\usepackage{comment}
\usepackage{amsmath}
\input{epsf}
\usepackage{psfrag}

\usepackage[usenames,dvipsnames]{color}

\newcommand{\beq}{\begin{eqnarray}}
\newcommand{\eeq}{\end{eqnarray}}
\newcommand{\Slash}[1]{{\ooalign{\hfil/\hfil\crcr$#1$}}}

\newcommand{\nn}{\nonumber \\}

\usepackage{amsmath,color}
\usepackage{amssymb}
\usepackage{bm}%
\usepackage{graphicx}
\usepackage{multirow}

\begin{document}

\title{Quantum tomography at the Electron-Ion Collider
}

\author{Michael Fucilla}
\affiliation{ National Centre for Nuclear Research, Pasteura 7, Warsaw 02-093, Poland}
\affiliation{Institute of Nuclear Physics, Polish Academy of Sciences, Radzikowskiego 152, 31-342 Cracow, Poland}

\author{Yuxun Guo}
\author{Yoshitaka Hatta  }
\affiliation{Physics Department, Brookhaven National Laboratory, Upton, NY 11973, USA}
\affiliation{RIKEN BNL Research Center, Brookhaven National Laboratory, Upton, NY 11973, USA}

\begin{abstract}
We initiate the study of quantum state tomography---the experimental reconstruction of the complete spin density matrix---of massless and massive quark-antiquark pairs produced at the  future Electron-Ion Collider  (EIC). We show that all  components of the spin density matrix can be accessed through  azimuthal angular correlations between hadron and dihadron pairs generated by the Collins and dihadron fragmentation functions. This enables the experimental investigation of a variety of quantum information properties at the EIC, such as entanglement and Bell-nonlocality. We predict that the recently discovered  maximal entanglement in the heavy-quark sector is signaled by a strong modification of   the usual Collins angular dependence $\cos(\phi+\phi')$.

\end{abstract}

\maketitle

\section{Introduction}

A notable  recent development at the LHC is the application of quantum state tomography to top-antitop ($t\bar{t}$) production. Shortly after the experimental establishment of quantum entanglement between the top and antitop spins \cite{ATLAS:2023fsd,CMS:2024pts}, the CMS collaboration \cite{CMS:2024zkc} has extracted the polarization vectors $B^a$, $\bar{B}^a$ and the complete set of spin-correlation coefficients $C^{ab}$ $(a,b=x,y,z)$ of the $t\bar{t}$ system
\beq
\rho = \frac{1}{4} \left(\mathbb{I}\otimes \mathbb{I}+B^a \sigma^a \otimes \mathbb{I}+\bar{B}^b\mathbb{I}\otimes \sigma^b+C^{ab}\sigma^a\otimes \sigma^b \right), \label{1}
\eeq
from the angular distributions of its weak decay products.
The resulting $4\times4$ density matrix $\rho$ provides a full characterization of the $t\bar t$ pairs as a two-qubit mixed quantum state, rather than merely measuring single-spin properties. The complete specification of the density matrix---quantum state tomography---opens the possibility of determining a variety of quantum information properties including entanglement \cite{Afik:2020onf,Dong:2023xiw,Han:2023fci}, Bell nonlocality \cite{Fabbrichesi:2021npl,Severi:2021cnj}, and quantum magic \cite{White:2024nuc,Li:2026zum} directly from collider data, turning the LHC into a laboratory for testing intricate quantum correlations among elementary particles at the highest available energies \cite{Barr:2024djo,Afik:2025ejh}.

In this paper, we initiate the discussion of quantum state tomography at the future Electron-Ion Collider (EIC) \cite{AbdulKhalek:2021gbh} at Brookhaven National Laboratory in the U.S. (see also \cite{Xiao:2026tbs}), focusing on the inclusive and exclusive productions of  quark-antiquark  pairs $u\bar{u}$, $d\bar{d}$, $s\bar{s}$ and $c\bar{c}$ in electron-proton ($ep$) and electron-nucleus ($eA$)  collisions.  The spin density matrix elements  $B,\bar{B},C$ of the pairs have been computed recently  \cite{Qi:2025onf,Fucilla:2025kit,Hatta:2025obw,Fucilla:2026mkg} (see also \cite{Cheng:2025zaw,Bloss:2026yrf,Hatta:2026dqs,Cao:2026zqo}), but so far  concrete observables for experimentally probing these matrix elements have not been clearly identified. A common strategy, also adopted by the CMS collaboration, is to exploit parity-violating weak decays \cite{PS185:2006yyx,BESIII:2018cnd,Galanti:2015pqa,Bernal:2023jba,CMS:2024zkc,STAR:2025njp,Lin:2025eci,Zhang:2026wvn,Kanachova:2026ywy}  to infer the spin correlation among  produced particles. While this can be pursued also at the EIC  \cite{Qi:2025onf,Fucilla:2025kit}, here we explore an approach based purely on the strong interaction. Our work is similar in spirit to  \cite{Cheng:2025cuv}, where the authors   applied the collinear dihadron fragmentation function (FF)  $H_1^\sphericalangle$ \cite{Bianconi:1999cd} to access the transverse  components of the $C$-matrix for quark-antiquark pairs in $e^+e^-$ annihilation. With the help of transverse momentum dependent (TMD) fragmentation functions (FFs), in particular the Collins FF $H_1^\perp$ \cite{Collins:1992kk} and the dihadron FF $G_1^\perp$ \cite{Bianconi:1999cd}, we demonstrate that all components of $B,\bar{B},C$ of the density matrix are accessible in the   more complicated environment  of electron-proton/nucleus collisions.

Conceptually, quantum state tomography is distinct from  conventional 3D tomography established as  one of the major scientific pillars of the EIC.  While the goal of 3D tomography is to map out the  multidimensional partonic structure of the  proton and nuclei in the initial state, quantum state tomography aims at the quantum mechanical  characterization of the final state. However, as shown in \cite{Fucilla:2025kit,Hatta:2026dqs,Fucilla:2026mkg}  and further demonstrated in this work,  the spin density matrix elements depend sensitively on the TMD parton distribution functions and the generalized parton distributions (GPDs) of the target. They thus constitute a new class of observables related to TMDs and GPDs, and  can be naturally integrated into the broader tomographic program   at the EIC, enriching its scope and strengthening its interdisciplinary character.

\section{New representation of the spin density matrix}

In this section, we derive useful representations for  the  spin density matrix elements of a massive quark-antiquark ($q\bar{q}$) pair produced in high energy scattering. We work in the center-of-mass (CM) frame of the pair and take the direction of the quark as the $z$-axis. The $x$ axis is chosen to be along the production plane spanned by the $q\bar{q}$ pair and initial colliding particles. The $y$-axis is then normal to the production plane.  The spin density matrix of the pair takes the general form (\ref{1}) where  $\mathbb{I}$ is the unit $2\times 2$ matrix and $\sigma^{a=x,y,z}$ are the Pauli matrices. In our convention, the matrix $C_{ab}$ represents  correlations between  the quark and antiquark spins, with the $z$-axis being their common quantization axis. This is different from correlations in helicity space: For the antiquark moving in the $-z$ direction, the spin $z$-component is opposite to its helicity.

Let us write generic $q\bar{q}$ production amplitudes as\footnote{Our conventions are
\beq
\epsilon^{0123}=\epsilon^{-+12}=\epsilon^{12}=1,\qquad \gamma^\mu = \begin{pmatrix} 0 & \sigma^\mu \\ \bar{\sigma}^\mu & 0 \end{pmatrix} \qquad \gamma_5=i\gamma^0\gamma^1\gamma^2\gamma^3={\rm diag}(-\mathbb{I},\mathbb{I}),
\eeq
where $\sigma^\mu=(1,\vec{\sigma})$ and $\bar{\sigma}^\mu=(1,-\vec{\sigma})$.
}
\beq
\bar{u}_\alpha(k)\Gamma v_{\alpha'
}(k'), \label{amp}
\eeq
where $\Gamma$ is a matrix in Dirac and color spaces. We parametrize the quark and antiquark momenta as
\beq
k^\mu = \frac{{\cal M}}{2}(1,0,0,\beta), \qquad k'^\mu =\frac{{\cal M}}{2}(1,0,0,-\beta),
\label{pp}
\eeq
where ${\cal M}^2=(k+k')^2$ is the invariant mass,  $\beta =\sqrt{1-\frac{4m^2}{{\cal M}^2}}$ is the velocity,    and $m$ is the quark mass.
The density matrix is obtained by
squaring (\ref{amp})  while keeping  the spin indices $\alpha,\alpha'$ different in the amplitude and the complex-conjugate amplitude
\beq
{\rm Tr}\left[\bar{\Gamma} u_\beta(k)\bar{u}_\alpha(k)\Gamma v_{\alpha'
}(k')\bar{v}_{\beta'}(k') \right]  = A\xi^\dagger_\alpha\eta^\dagger_{\alpha'} \rho\, \xi_\beta \eta_{\beta'} , \label{trace}
\eeq
where we defined $\bar{\Gamma}\equiv \gamma^0\Gamma^\dagger \gamma^0$. It is  assumed that all Lorentz and color indices in $\Gamma$ have been traced over. $\xi$ and $\eta$ are the two-component spinors initially contained in the four-component Dirac spinors $u$ and $v$.  The computation of $\rho$ thus requires the reduction of the four-dimensional Dirac space to the two-dimensional spinor space. While this is  straightforward and has been routinely performed in the literature \cite{Bernreuther:1993hq,Chen:1994ar}, here we present formulas expressing $\vec{B},\vec{B}'$ and $C$ directly in terms of four-dimensional quantities ($i,j=x,y$)
\beq
C^{ij}&=& \frac{{\rm Tr}\left[\bar{\Gamma} \gamma^i\gamma_5(\Slash k+m)\Gamma \gamma^j\gamma_5(\Slash k'-m)\right] }{ {\rm Tr} \left[\bar{\Gamma}(\Slash k+m) \Gamma (\Slash k'-m)\right]},\label{full}
\nn
C^{zi}&=& \frac{{\rm Tr}\left[\bar{\Gamma}\gamma_5\gamma^0\gamma^3(\Slash k+m)\Gamma \gamma^i\gamma_5(\Slash k'-m)\right] }{ {\rm Tr} \left[\bar{\Gamma} (\Slash k+m) \Gamma (\Slash k'-m)\right]}, \nn
C^{iz}&=&  -\frac{{\rm Tr}\left[\bar{\Gamma} \gamma^i\gamma_5(\Slash k+m)\Gamma \gamma_5\gamma^0\gamma^3(\Slash k'-m)\right] }{ {\rm Tr} \left[\bar{\Gamma} (\Slash k+m) \Gamma (\Slash k'-m)\right]}, \nn
C^{zz}&=&-\frac{{\rm Tr}\left[\bar{\Gamma}\gamma_5\gamma^0\gamma^3(\Slash k+m)\Gamma \gamma_5\gamma^0\gamma^3(\Slash k'-m)\right] }{ {\rm Tr} \left[\bar{\Gamma}(\Slash k+m) \Gamma (\Slash k'-m)\right]} . \label{cij}
\eeq
\beq
B^i=\frac{{\rm Tr}\left[\bar{\Gamma}\gamma^i\gamma_5(\Slash k+m)\Gamma (\Slash k'-m)\right] }{ {\rm Tr} \left[\bar{\Gamma} (\Slash k+m) \Gamma (\Slash k'-m)\right]}, \qquad  B^z= \frac{{\rm Tr}\left[\bar{\Gamma}\gamma_5\gamma^0\gamma^3(\Slash k+m)\Gamma (\Slash k'-m)\right] }{ {\rm Tr} \left[\bar{\Gamma} (\Slash k+m) \Gamma (\Slash k'-m)\right]}, \nn
\bar{B}^i=\frac{{\rm Tr}\left[\bar{\Gamma}(\Slash k+m)\Gamma \gamma^i\gamma_5 (\Slash k'-m)\right] }{ {\rm Tr} \left[\bar{\Gamma} (\Slash k+m) \Gamma (\Slash k'-m)\right]},\qquad \bar{B}^z=-\frac{{\rm Tr}\left[\bar{\Gamma}(\Slash k+m)\Gamma \gamma_5\gamma^0\gamma^3(\Slash k'-m)\right] }{ {\rm Tr} \left[\bar{\Gamma} (\Slash k+m) \Gamma (\Slash k'-m)\right]} . \label{bb}
\eeq
The derivation is given in Appendix \ref{appA}. If the underlying production mechanism conserves parity, $B^{x}=B^z=\bar{B}^x=\bar{B}^z=0$ and $C^{xy}=C^{yx}=C^{yz}=C^{zy}=0$.

While the formulas (\ref{cij}), (\ref{bb}) may seem somewhat familiar, to our knowledge they have not previously been  written down explicitly in this form. Two quick comments are in order. First, since $k^\mu$, $k'^\mu$  have only the $\mu=0,z$ components, the ordering of $\Slash k+m$ can be changed as
\beq
\gamma_5\gamma^0\gamma^3(\Slash k+m)=(\Slash k+m) \gamma_5\gamma^0\gamma^3 ,\qquad \gamma^i\gamma_5(\Slash k+m) =(\Slash k+m) \gamma^i\gamma_5, \label{lr}
\eeq
and similarly for $\Slash k'-m$.
(Note that $\gamma_5\gamma^0\gamma^3=i\gamma^1\gamma^2$.)
Second, the  formulas simplify for a massless quark
\beq
\gamma_5\gamma^0\gamma^3(\Slash k+m)\to  \gamma_5\Slash k, \qquad \gamma_5\gamma^0\gamma^3 (\Slash k'-m)\to -\gamma_5\Slash k', \label{gamma5}
\eeq
\beq
\gamma^i\gamma_5(\Slash k+m) \to  i\sigma^{i-}\gamma_5 k^+, \qquad \gamma^j\gamma_5(\Slash k'-m)\to i\sigma^{j+}\gamma_5k'^- .\label{massless}
\eeq
These  $\gamma$-matrix structures are commonly associated with longitudinally and transversely polarized (anti)quarks.

\section{TMD fragmentation function }

The representation (\ref{cij}) is particularly useful when connecting to fragmentation functions (FFs) which are usually classified in terms of independent $\gamma$-matrices. In this section we provide a minimal account of transverse momentum dependent (TMD) FFs mainly to set up the notations to be used in the later sections. See \cite{Metz:2016swz} for a comprehensive review.

For a quark moving in the $+z$ direction and fragmenting into a hadron with momentum $P_h$ carrying a momentum fraction $z$, TMD FFs are defined by\footnote{Most references show FFs of a quark moving in the $-z$ direction.
\beq
\Delta^{h/q}(z,P_{h\perp}) &=& \frac{1}{z} \int \frac{ dx^+d^2x_\perp}{(2\pi)^3} e^{i\frac{P_h^-}{z} x^+  -ik_\perp\cdot x_\perp}  \sum_X\langle 0|\psi(x^+,x_\perp)|P_h X\rangle \langle P_h X|\bar{\psi}(0)|0\rangle\Big\arrowvert_{P_{h\perp}=0}   \nn
&=& \gamma^+D^{h/q}_{1}(z,z^2k_\perp^2)
-i\sigma^{i+}\gamma_5 %
\frac{\epsilon^{ij}k_\perp^j}{M}H_{1}^{\perp h/q}(z,z^2k_\perp^2).
\eeq
This is obtained from  (\ref{ffsingle}) by swapping $+\leftrightarrow -$. A caveat is that the epsilon tensor leads to a sign flip since $\epsilon^{ij}=\epsilon^{-+ij}=-\epsilon^{+-ij}$.
}
\beq
\Delta^{h/q}(z,P_{h\perp}) &=& \frac{1}{z} \int \frac{ dx^-d^2x_\perp}{(2\pi)^3} e^{i\frac{P_h^+}{z} x^-  -ik_\perp\cdot x_\perp}  \sum_X\langle 0|\psi(x^-,x_\perp)|P_h X\rangle \langle P_h X|\bar{\psi}(0)|0\rangle\Big\arrowvert_{P_{h\perp}=0}   \nn
&=& \gamma^-D^{h/q}_{1}(z,z^2k_\perp^2)
+i\sigma^{i-}\gamma_5 %
\frac{\epsilon^{ij}k_\perp^j}{M}H_{1}^{\perp h/q}(z,z^2k_\perp^2)+\cdots \; ,
\label{ffsingle}
\eeq
with $k^+=\frac{P^+_h}{z}$, $k_\perp = - \frac{P_{h\perp}}{z}$ so that $z^2k_\perp^2=P_{h\perp}^2$. $D_1$ is the unpolarized TMD FF related to the collinear FF as
\beq
\int d^2P_{h\perp} D_1^{h/q}(z,z^2k_\perp^2)=D^{h/q}_1(z) \; .
\eeq
$H_1^\perp$ is the Collins FF representing the fragmentation of a transversely polarized quark \cite{Collins:1992kk}.
(\ref{ffsingle}) is defined in a frame in which $P_{h\perp}=0$
and the fragmenting quark has nonzero transverse momentum $k_\perp\neq 0$. One can perform a transverse boost to a frame in which the fragmenting quark has vanishing transverse momentum. The hadron then has $P_{h\perp}=-zk_\perp$ in this frame.
For a left-moving antiquark,
\beq
\Delta^{h/\bar{q}}(z,P_{h\perp}) &=& \frac{1}{z} \int \frac{ dx^+d^2x_\perp}{(2\pi)^3}e^{i\frac{P^-_h}{z} x^+  -ik_\perp\cdot x_\perp}    \sum_X\langle 0|\bar{\psi}(x^+,x_\perp)|P_h X\rangle \langle P_h X|\psi(0)|0\rangle\Big\arrowvert_{P_{h\perp}=0}  \nn
&=& \gamma^+D_{1}^{h/\bar{q}}(z,z^2k_\perp^2)
-\frac{i\sigma^{i+}\gamma_5\epsilon^{ij}k^j_{\perp} }{M}H_{1}^{\perp h/\bar{q}}(z,z^2k_\perp^2) +\cdots.
\eeq

If the hadron $h$ has spin (for example the $\Lambda$ baryon), there are spin-dependent TMD FFs to be added to the right hand side of (\ref{ffsingle}). Their collinear ($P_{h\perp}$-integrated) versions have been studied in the context of final state polarimetry and entanglement  \cite{Galanti:2015pqa,Kats:2023zxb,Lin:2025eci}. However, in this paper we neglect the spin of $h$, or assume $h$ is spinless.
The other terms neglected in (\ref{ffsingle}) are higher twist FFs which in particular include the quark mass effect. They can be partly restored by applying the operation (\ref{massless}) in reverse
\beq
k^+\Delta^{h/q}&\to&  (\Slash k+m)D_{1}^{h/q}(z,z^2k_\perp^2)
+\gamma^i\gamma_5(\Slash k+m)
\frac{\epsilon^{ij}k_\perp^j}{M}H_{1}^{\perp h/q}(z,z^2k_\perp^2), \label{q1}
\eeq
\beq
k'^-\Delta^{h/\bar{q}} &\to& (\Slash k'-m)D_{1}^{h/\bar{q}}(z,z^2k_\perp^2)
-\gamma^i\gamma_5(\Slash k'-m)
 \frac{\epsilon^{ij}k_\perp^j}{M}H_{1}^{\perp h/\bar{q}}(z,z^2k_\perp^2) .\label{q2}
\eeq
This amounts to taking into account twist-3 and twist-4 FFs
\beq
k^+\Delta = k^+\gamma^-D_1(z)+ME_3(z)+ \frac{M^2}{k^+}\gamma^+F_4(z)+\cdots,
\eeq
and only keeping the kinematical part related to the leading twist FF  \cite{Boer:2008fr}
\beq
&&E_3(z)=\frac{m}{M}D_1(z)+\tilde{E}_3(z) ,
\qquad  F_4(z)=\frac{m^2}{2M^2}D_1(z)+\tilde{F}_4(z),
\label{34}
\eeq
while neglecting the dynamical (`genuine higher-twist') parts $\tilde{E}_3$ and $\tilde{F}_4$. Admittedly, this may not be a consistent truncation of the higher  twist contributions, but  rather a pragmatic prescription for  enforcing the equation of motion relations $(\Slash k-m)\Delta^{h/q}=\Delta^{h/q}(\Slash k-m)=0$ and $(\Slash k'+m)\Delta^{h/\bar{q}}=\Delta^{h/\bar{q}}(\Slash k'+m)=0$. Nevertheless, it appears to have  intrinsic value, as it  enables a smooth connection to the spin density matrix formalism. In the following, we adopt (\ref{q1}) and (\ref{q2}) as a  minimal prescription for incorporating quark mass effects, while leaving a more rigorous  justification for future work.

The Collins FF is  sensitive only to the transverse polarization of the quark and antiquark. In order to access  the longitudinal polarization, we need to introduce the dihadron FFs  describing the process $q\to h_1(P_1)+h_2(P_2)+X$ \cite{Bianconi:1999cd,Metz:2016swz}. For a quark moving in the $+z$ direction, they are defined as
\beq
\Delta^{h_1h_2/q}(z,\zeta,k_\perp,P_h, R) &=& \frac{1}{z} \int \frac{ dx^-d^2x_\perp}{(2\pi)^3} e^{ik^+ x^-  -ik_\perp\cdot x_\perp}  \sum_X\langle 0|\psi(x^-,x_\perp)|P_1P_2 X\rangle \langle P_1P_2 X|\bar{\psi}(0)|0\rangle\Big\arrowvert_{P_{h\perp}=0}   \nn
&=& \gamma^-D_1^{h_1h_2/q}(z,\zeta,P_h,R) +    \gamma^-\gamma_5\frac{\epsilon^{ij}R^i_\perp k_\perp^j}{M_1M_2}G_1^{\perp h_1h_2/q}(z,\zeta,P_h,R) \\
&& +i\sigma^{i-}\gamma_5\left(H_1^{\sphericalangle h_1h_2/q} (z,\zeta,P_h,R) \frac{\epsilon^{ij}R^j_\perp}{M_1+M_2}  +H_1^{\perp h_1h_2/q}(z,\zeta,P_h,R) \frac{\epsilon^{ij}k_\perp^j}{M_1+M_2}\right), \notag
\label{diFF}
\eeq
where $P_h=P_1+P_2$,  $R=\frac{P_{1}-P_{2}}{2}$ and
\beq
k^+=\frac{P_h^+}{z},\qquad k_\perp = -\frac{P_{h\perp}}{z} ,
\eeq
\beq
P_1^+=\frac{1+\zeta}{2}P_h^+, \qquad P_2^+=\frac{1-\zeta}{2}P_h^+.
\eeq
The hadron pair's invariant mass ${\cal M}_h^2=(P_1+P_2)^2$ is related to $R_\perp$ as
\beq
R_\perp^2=\frac{1-\zeta^2}{4}{\cal M}^2_h-\frac{1-\zeta}{2}M_1^2-\frac{1+\zeta}{2}M_2^2.
\eeq
$H_1^\perp$ is the dihadron analog of the Collins FF, whereas  $G_1^\perp$ and $H_1^\sphericalangle$ represent genuinely new effects.
Phenomenologically, the collinear version of $H_1^\sphericalangle$
\beq
&& H_1^{\sphericalangle}(z,\zeta,{\cal M}^2_h)=\int d^2P_{h\perp} \left(H_1^\sphericalangle(z,\zeta,P_h,R)+\frac{\vec{k}_\perp \cdot \vec{R}_\perp}{R_\perp^2}H_1^\perp(z,\zeta,P_h,R)\right), \label{hcoll}
\eeq
sometimes referred to as  the `interference' fragmentation function,
has been most well-studied \cite{Boer:2003ya,Bacchetta:2008wb,Courtoy:2012ry,Cocuzza:2023vqs,Cocuzza:2023oam}.  Our main interest here is rather
$G_1^\perp$, which allows us to access the longitudinal polarization of the quark even when $h_1,h_2$ are spinless. Again, we minimally include the quark mass effect
\beq
k^+\Delta^{h_1h_2/q} &\to&
(\Slash k+m)D_1^{h_1h_2/q}(z,\zeta,P_h,R)-\gamma_5\gamma^0\gamma^3(\Slash k+m)\frac{\epsilon^{ij}R^i_\perp k_\perp^j}{M_1M_2}G_1^{\perp h_1h_2/q}(z,\zeta,P_h,R) \nn
&& +\gamma^i\gamma_5(\Slash k+m)\left(H_1^{\sphericalangle h_1h_2/q} (z,\zeta,P_h,R) \frac{\epsilon^{ij}R^j_\perp}{M_1+M_2}  +H_1^{\perp h_1h_2/q}(z,\zeta,P_h,R)  \frac{\epsilon^{ij}k_\perp^j}{M_1+M_2}\right) \; . \nonumber \\
\eeq
For an antiquark moving in the $-z$ direction, we use
\beq
&& k'^-\Delta^{h_1h_2/\bar{q}} \to
(\Slash k'-m)D_1^{h_1h_2/\bar{q}}(z,\zeta,P_h,R)+\gamma_5\gamma^0\gamma^3(\Slash k'-m)\frac{\epsilon^{ij}R^i_\perp k'^j_\perp}{M_1M_2}G_1^{\perp h_1h_2/\bar{q}}(z,\zeta,P_h,R) \nn
&& \qquad \quad -\gamma^i\gamma_5(\Slash k'-m)\left(H_1^{\sphericalangle h_1h_2/\bar{q}}(z,\zeta,P_h,R) \frac{\epsilon^{ij}R^j_\perp}{M_1+M_2}  +H_1^{\perp h_1h_2/\bar{q}}(z,\zeta,P_h,R) \frac{\epsilon^{ij}k'^j_\perp}{M_1+M_2}\right).
\eeq
The relative sign in front of $G_1^{\perp h_1h_2/\bar{q}}$ is due to a minus sign from switching directions (see Footnote 2), a minus sign associated with the $C$-parity transform of $\gamma^\mu\gamma_5$ \cite{Boer:1997mf}, and another  minus sign in
(\ref{gamma5}).

\section{$e^+e^-$ annihilation}

We are now ready to discuss the relation between the spin density matrix elements and angular correlations among back-to-back hadrons detected  in the fragments of the quark and the antiquark. We start with the simpler case of $e^+e^-$ annihilation and consider both hadron pair and dihadron pair productions
\beq
e^-(\ell)+e^+(\ell')\to \gamma^*\to q(k)+\bar{q}(k')\to \begin{cases} h(P)+h'(P')+X, \\
h_1(P_1)+h_2(P_2)+h'_1(P'_1)+h'_2(P'_2)+X, \end{cases}
\label{qqhh}
\eeq
ignoring the $Z$-boson or other resonance  contributions.
It should be mentioned that there is already an extensive literature on this topic \cite{Artru:1995zu,Boer:1997mf,Boer:2003ya,Anselmino:2007fs,Bacchetta:2008wb,Pitonyak:2013dsu,Metz:2016swz}. Our sole intention here is to reformulate the discussion in terms of the spin density matrix with an emphasis on  the quark mass effect.

In unpolarized $e^+e^-$ annihilation into a massive $q\bar{q}$ pair, the spin density matrix of the produced $q\bar{q}$ pair  is given by \cite{Bernreuther:1993hq}
\beq
C^{ab}=\frac{1}{1+\cos^2\theta+\frac{4m^2}{s}\sin^2\theta}\begin{pmatrix}\left(1+\frac{4m^2}{s}\right)\sin^2\theta  & 0 & -\frac{2m}{\sqrt{s}}\sin 2\theta \\ 0 &  -\left(1-\frac{4m^2} {s}\right)\sin^2\theta &  0 \\ -\frac{2m}{\sqrt{s}}\sin 2\theta & 0 &
1+\cos^2\theta- \frac{4m^2}{s}\sin^2\theta \end{pmatrix} \; ,
\label{xyz}
\eeq
where $a,b=x,y,z$ and $s=(\ell+\ell')^2$ is the CM energy. $B^a=\bar{B}^a=0$ at tree level. The $z$ axis is taken along the quark direction, and the event plane is identified with the $xz$ plane.  $\theta$ is the angle between the incoming electron and the outgoing quark in this plane.

In the literature, two  different choices for the $z$-axis  have been used  to describe  azimuthal asymmetries of produced hadrons \cite{Metz:2016swz}. In the asymmetric frame choice, the $z$-axis is taken along   the momentum of one of the hadrons \cite{Boer:1997mf} (or $\vec{P}'_{h}=\vec{P}'_1+\vec{P}'_2$ in the dihadron case \cite{Boer:2003ya}). In the symmetric frame choice, the thrust axis   defines the $z$-axis \cite{Anselmino:2007fs,Boer:2008fr}. From the viewpoint of TMD factorization,  the first choice is preferred \cite{Boussarie:2023izj}. Here, however, we mostly use the second choice, as it provides a common reference frame for both the  single-hadron and dihadron analyses. Moreover, the symmetric choice appears to be more straightforwardly generalizable  to the DIS problem to be discussed in the next section. Experimentalists often present asymmetries measured in both frames   \cite{Belle:2008fdv,BaBar:2013jdt}.

Let us first consider single-hadron FFs. To leading order, the thrust axis coincides with the quark momentum $\vec{k}=-\vec{k}'$.
For massless quark, the parton-level cross section is given by
\beq
d\sigma \propto {\rm Tr}[\gamma_\mu \Slash k' \gamma_\nu \Slash k](\ell^\mu \ell'^\nu+\ell^\nu \ell'^\mu -g^{\mu\nu}\ell\cdot \ell') \; ,
\eeq
where $\ell^\mu=\frac{\sqrt{s}}{2}(1,-\sin\theta,0,\cos\theta)$, $\ell'^\mu=\frac{\sqrt{s}}{2}(1,\sin\theta, 0,-\cos\theta)$,  $k^\mu=\frac{\sqrt{s}}{2}(1,0,0,1)$ and $k'^\mu=\frac{\sqrt{s}}{2}(1,0,0,-1)$.
The hadron-level cross section is obtained by  substituting
\beq
\Slash k \to \Slash k D_1
+i\sigma^{i-}\gamma_5 k^+
\frac{\epsilon^{ij}k_\perp^j}{M}H_{1}^{\perp}, \qquad \Slash k' \to \Slash k' \bar{D}_1
-i\sigma^{i+}\gamma_5k'^-\frac{\epsilon^{ij}k'^j_{\perp} }{M'}\bar{H}_1^\perp.
\eeq
This leads to the formula
\beq
\frac{d\sigma}{d\cos\theta dz dz' d^2P_\perp d^2P'_\perp} = \frac{N_c\pi \alpha^2}{2s} \sum_q e_q^2\Biggl\{(1+\cos^2\theta)  D_1^{h/q}(z,z^2k^2_\perp)D_1^{h'/\bar{q}}(z',z'^2k'^2_\perp) \nn
+\sin^2\theta  \cos(\phi_P+\phi_{P'}) \frac{P_\perp P'_\perp H_1^{\perp h/q }(z,z^2k^2_\perp)H_1^{\perp h/\bar{q}}(z',z'^2k'^2_\perp)}{zz'MM'}  + (q\leftrightarrow \bar{q})\Biggr\} \nn = \frac{N_c\pi \alpha^2}{2s}(1+\cos^2\theta)   \sum_q e_q^2\Biggl\{ D_1\bar{D}_1
- \epsilon^{il} P^l_\perp \epsilon^{jm}P'^m_\perp C_0^{ij} \frac{H_1^\perp \bar{H}_1^{\perp }}{zz'MM'} \Biggr\}.
\eeq
The $\cos(\phi_P+\phi_{P'})$ azimuthal modulation is known as Collins asymmetry.
The second line (with $i,j=x,y$) is written in terms of the spin density matrix elements $C_0^{ij}$  (\ref{xyz}) with $m=0$.

For massive quarks we use (\ref{q1}) and (\ref{q2}) instead and find, for a single quark flavor,
 \beq
 && \frac{d\sigma_q}{d\cos\theta dz dz' d^2P_\perp d^2P'_\perp} = \frac{N_c\pi \alpha^2}{2s}\left(1+\cos^2\theta+\frac{4m_q^2}{s}\sin^2\theta\right)   e_q^2 \beta_q \nn && \qquad \times \Biggl\{ D_1\bar{D}_1
-B^i \epsilon^{ij}P^j\frac{H_1^\perp \bar{D}_1}{zM}  + \bar{B}^i\epsilon^{ij}P'^j   \frac{D_1 \bar{H}_1^{\perp }}{z'M'}    - \epsilon^{il} P^l_\perp \epsilon^{jm}P'^m_\perp C^{ij} \frac{H_1^{\perp} \bar{H}_1^{\perp }}{zz'MM'} \Biggr\} \; ,
 \eeq
where the factor $ \beta_q \equiv \sqrt{1-4m_q^2/s}$ comes from the two-body phase space for massive particles and $C^{ij}$ is the full matrix (\ref{xyz}) with the corresponding quark mass $m_q$. We have included the polarization terms, although $B^i=\bar{B}^i=0$ to leading  order. The spin correlation term features a new azimuthal modulation $\cos(\phi_P-\phi_{P'})$ absent in the massless theory
\beq
- \epsilon^{il} P^l_\perp \epsilon^{jm}P'^m_\perp C^{ij} = P_\perp P'_\perp \frac{\sin^2\theta}{1+ \cos^2 \theta + \frac{4m_q^2}{s} \sin^2 \theta}\left(\cos(\phi_P+\phi_{P'}) -\frac{4m_q^2}{s}\cos(\phi_P-\phi_{P'}) \right).
\eeq
If one is only interested in extracting the coefficients $C^{ij}$, one can integrate over the magnitudes of  $P_\perp$ and  $P'_\perp$
\beq
&&
\frac{zz' d\sigma_q}{d\cos\theta dz dz' d\phi_P d\phi_{P'}} = \frac{N_c\pi \alpha^2}{2s}\left(1+\cos^2\theta+\frac{4m^2_q}{s}\sin^2\theta\right)   \frac{e_q^2\beta_q }{4\pi^2}
\Biggl\{ z D_1(z) z' \bar{D}_1(z') \label{h1h1} \\
&& \qquad -B^i\epsilon^{ij}\hat{P}^j_\perp H_1^{\perp(1)}(z)z'\bar{D}_1(z')+\bar{B}^i\epsilon^{ij}\hat{P}'^j_\perp zD_1(z)\bar{H}_1^{\perp(1)}(z')
- \epsilon^{il} \hat{P}^l_\perp \epsilon^{jm} \hat{P}'^m_\perp C^{ij} H_1^{\perp(1)}(z) \bar{H}_1^{\perp(1)}(z')\Biggr\} ,
\notag
 \eeq
where $\hat{P}_\perp=(\cos\phi_P,\sin\phi_P)$, $\hat{P}'_\perp=(\cos\phi_{P'},\sin\phi_{P'})$  and we defined
\beq
H_1^{\perp(1)}(z)
= \int d^2P_\perp \frac{| \vec{P}_\perp|}{M}  H_1^{\perp h/q}(z,z^2P_\perp^2). \label{half}
\eeq
One may further integrate over $z,z'$ and parametrize the result in terms of the momentum fraction $\langle z\rangle_{h/q}=\int_0^1 dz z D_1^{h/q}(z)$. (Note  the momentum sum fule  $\sum_h \langle z\rangle_{h/q}=1$.) However, it is advantageous to stay differential in $z,z'$, since asymmetries are typically larger in the large-$z,z'$ region, see the numerical section.

The Collins FF gives access only to the transverse  components $C^{ij}$, $i,j=x,y$. To probe the full density matrix $C^{ab}$, $a,b=x,y,z$,  we need  the $G_1^\perp$ dihadron FF.
The cross section takes the form
(see \cite{Boer:2003ya,Matevosyan:2018icf} for a complete treatment in the asymmetric frame)
\beq
&& \frac{d\sigma_q}{d\cos\theta dz dz'd\zeta d\zeta' d{\cal M}_h^2 d{\cal M}_{h'}^2 d\phi_R d\phi_{R'} d^2P_{h\perp} d^2P'_{h\perp} } =\frac{N_c\pi \alpha^2}{2s}\left(1+\cos^2\theta+\frac{4m^2_q}{s}\sin^2\theta\right) e_q^2 \beta_q \nn
&& \times \Biggl[D_1^{h_1h_2/q}(z,\zeta,P_h,R)D_1^{h'_1h'_2/\bar{q}}(z',\zeta',P'_h,R')  \nn && \qquad + \bar{B}^zD_1 \frac{R'_\perp P'_{h\perp}}{z'M'_1M'_2}\bar{G}_1^\perp\sin(\phi_{R'}-\phi_{P'_h})+ B^z\frac{R_\perp P_{h\perp}}{zM_1M_2}G_1^\perp\bar{D}_1  \sin(\phi_{R}-\phi_{P_h})   \nn
&& \qquad +C^{zz}\frac{R_\perp R'_\perp P_{h\perp}P'_{h\perp}}{zz'M_1M_2M'_1M'_2} G_1^\perp \bar{G}_1^{\perp} \sin(\phi_R-\phi_{P_h})\sin(\phi_{R'}-\phi_{P'_h}) \nn
&& \qquad  - C^{iz} \epsilon^{ij} \left(\frac{R_\perp^j H_1^{\sphericalangle}}{M_1+M_2} -\frac{P_{h\perp}^j H_1^{\perp}}{z(M_1+M_2)}\right)  \frac{\sin(\phi_{R'}-\phi_{P'_h})R'_\perp P'_{h\perp}}{z'M'_1M'_2}  \bar{G}_1^{\perp}\nn && \qquad +\frac{\sin(\phi_R-\phi_{P_h})R_\perp P_{h\perp}}{zM_1M_2}  G_1^{\perp}  C^{zi}\epsilon^{ij}\left(\frac{R'^j_\perp \bar{H}_1^{\sphericalangle}}{M'_1+M'_2}-\frac{P'^j_{h\perp} \bar{H}_1^{\perp}}{z'(M'_1+M'_2)}\right) \nn
&& \qquad  - C^{ij}\epsilon^{il}\left(\frac{R^l_\perp H_1^\sphericalangle}{M_1+M_2}-\frac{P^l_{h\perp} H_1^\perp}{z(M_1+M_2)}\right) \epsilon^{jm}\left(\frac{R'^m_\perp \bar{H}_1^{\sphericalangle}}{M'_1+M'_2}-\frac{P'^m_{h\perp} \bar{H}_1^{\perp}}{z'(M'_1+M'_2)}\right)
 +\cdots\Biggr]. \label{most}
\eeq
In order to extract the $C^{zz}$ component, the relative angles  $\phi_{RP}=\phi_R-\phi_{P_h}$ and $\phi'_{RP}=\phi_{R'}-\phi_{P'_h}$ need to be measured (see also \cite{Matevosyan:2017liq}). Integrating over the other variables,  we find
\beq
&& \frac{zz' d\sigma_q}{d\cos\theta dz dz'd\phi_{RP} d\phi'_{RP} } =\frac{N_c\pi \alpha^2}{2s}\left(1+\cos^2\theta+\frac{4m^2_q}{s}\sin^2\theta\right)  \nn
&& \qquad \times \frac{e_q^2\beta_q } {(2\pi)^2} \Biggl[zD_1^{(0)}(z)z'\bar{D}_1^{(0)}(z')+C^{zz}G_1^{\perp(1)}(z)\bar{G}_1^{\perp(1)}(z')  \sin(\phi_{RP})\sin(\phi'_{RP}) \Biggr],
\eeq
where
\beq
&& D_1^{h_1h_2/q(0)}(z)=\int d\zeta d{\cal M}_h^2\int d^2P_{h_\perp}   D_1^{h_1h_2/q}(z,\zeta,P_h,R), \nn
&& G_1^{\perp h_1h_2/q(1)}(z)=\int d\zeta d{\cal M}_h^2\int d^2P_{h\perp}  \frac{|\vec{P}_{h\perp}||\vec{R}_\perp|}{M_1M_2} G_1^{\perp h_1h_2/q}(z,\zeta,P_h,R).
\eeq
On the other hand, for the extraction of $C^{iz}$ (or $C^{zi}$), it is not necessary to measure the angle of  $P_{h\perp}$ (or $P'_{h\perp}$)
\beq
&& \frac{zz'd\sigma_q}{d\cos\theta dz dz'd\phi_R d\phi'_{RP}}=\frac{N_c\pi \alpha^2}{2s}\left(1+\cos^2\theta+\frac{4m^2_q}{s}\sin^2\theta\right) \nn
&& \qquad  \times \frac{e_q^2\beta_q}{(2\pi)^2}  \left[zD_1^{(0)}(z) z' \bar{D}^{(0)}_1(z')-C^{iz}\epsilon^{ij}\hat{R}^j_\perp H_1^{\sphericalangle(1)}(z) \sin \phi'_{RP}\bar{G}_1^{\perp (1)}(z')\right], \label{off}
\eeq
where
\beq
H_1^{\sphericalangle h_1h_2/q(1)}(z)=z\int d\zeta d{\cal M}_h^2  \frac{|\vec{R}_\perp|}{M_1+M_2}H_1^{\sphericalangle h_1h_2/q}(z,\zeta,{\cal M}_h^2) \; .
\eeq
 Finally, to access $C^{ij}$, one can integrate over both $d^2P_{h\perp}$ and $d^2P'_{h\perp}$.
 Only the last line $\propto H_1^\sphericalangle \bar{H}_1^\sphericalangle  \cos(\phi_R+\phi_{R'})$ survives, known as the Artru-Collins asymmetry  \cite{Artru:1995zu}. This   was exploited in  \cite{Cheng:2025cuv} for the purpose of testing Bell-nonlocality. We however  note that, as far as the components $C^{ij}$ are concerned,  the Collins asymmetry  (\ref{h1h1}) provides an alternative  method, reducing the complication of measuring dihadron pairs to hadron pairs.

The above results have been obtained in the symmetric (thrust axis) frame. For a later purpose, let us derive the generalization of the Collins asymmetry in the asymmetric frame \cite{Boer:1997mf}
\beq
\frac{d\sigma}{d\phi_Pdz dz'} &\sim&   \int d^2k_\perp d^2k'_\perp |\vec{k}_\perp|H_1^\perp(z,k^2_\perp) |\vec{k}'_\perp|H_1^\perp(z',k'^2_\perp) \delta^{(2)}\left(\vec{k}_\perp+\vec{k}'_\perp+\vec{P}_\perp/z\right)\nn && \times \cos (2\phi_P) \frac{\sin^2\theta}{1+\cos^2\theta}\cos(\phi_k+\phi_{k'}) , \label{eee}
\eeq
where $\phi_P$ is the angle of $h$ measured with respect to the plane spanned by the incoming leptons and $h'$. $\Delta \phi_k,\Delta \phi_{k'}$ are the angles of $\vec{k}_\perp,\vec{k}'_\perp$ relative to $\vec{P}_\perp$. In terms of the spin density matrix, (\ref{eee}) can be written as\footnote{There is an extra term $(C^{xy}-C^{yx})\sin(\Delta\phi_k-\Delta \phi_{k'})$ in the second line, but it  vanishes after integrating over the angles  $\phi_k,\phi_{k'}$. This means that the $C^{xy},C^{yx}$ components (although they are zero in parity-conserving QCD) cannot be probed experimentally if one works in the asymmetric frame.}
\beq
\frac{d\sigma}{d\phi_P dz dz'} &\sim&   \int d^2k_\perp d^2k'_\perp |\vec{k}_\perp|H_1^\perp(z,k^2_\perp) |\vec{k}'_\perp|H_1^\perp(z',k'^2_\perp) \delta^{(2)}\left(\vec{k}_\perp+\vec{k}'_\perp+\vec{P}_\perp/z\right) \nn
&& \times\left( \cos (2\phi_P) \frac{C^{xx}-C^{yy}}{2}\cos(\Delta\phi_k+\Delta\phi_{k'}) -\frac{C^{xx}+C^{yy}}{2} \cos (\Delta \phi_k-\Delta \phi_{k'})\right) . \label{mod}
\eeq
This formula can be used for massive quark production in $e^+e^-$ annihilation.
Notice that the added term  does not depend on the angle  $\phi_P$. It acts as the `pedestal' part of the $\phi_P$-distribution that often complicates the extraction of the $\cos(2\phi_P)$ modulation.

\section{Deep Inelastic Scattering}

In  $e^+e^-$ annihilation studied in the previous section, the $C$-matrix   (\ref{xyz}) follows from a simple leading order calculation and  depends only on basic kinematical variables. The coefficient $\frac{\sin^2\theta}{1+\cos^2\theta}$ is so standard  that one does not usually interpret it as a spin density matrix element. Rather, it is simply an input for extracting the Collins and dihadron FFs.
The situation is  different for massive $q\bar{q}$ production in electron-proton ($ep$) or electron-nucleus ($eA$) scattering which we now turn to. In this case, the $C$-matrix tends to be   significantly more complicated even at leading order \cite{Qi:2025onf,Fucilla:2025kit,Hatta:2025obw,Fucilla:2026mkg}, although it simplifies dramatically in the massless case. Moreover, beyond the one-gluon exchange studied in \cite{Qi:2025onf}, the $C$-matrix becomes sensitive to the nonperturbative multi-dimensional structure of the target proton/nucleus \cite{Fucilla:2025kit,Hatta:2025obw,Fucilla:2026mkg}. In inclusive $q\bar{q}$ production,  it can depend on the transverse momentum dependent distributions (TMD) of the target.  In exclusive  production, it depends on the generalized parton distributions (GPD) \cite{Hatta:2025obw}, or the generalized transverse momentum dependent distributions (GTMD) \cite{Fucilla:2025kit}, depending on the kinematics.
It has been demonstrated that such nonperturbative effects significantly modify and enrich the pattern of entanglement relative to the leading order/massless result. This brings us to the concept of quantum state tomography  in $ep$ and $eA$ collisions. The $C$-matrix element can be experimentally extracted, this time using the Collins and dihadron FFs as inputs. The result not only reveals the entanglement property of the $q\bar{q}$ pair, but also probes the nonperturbative multidimensional structure of the target proton/nucleus.

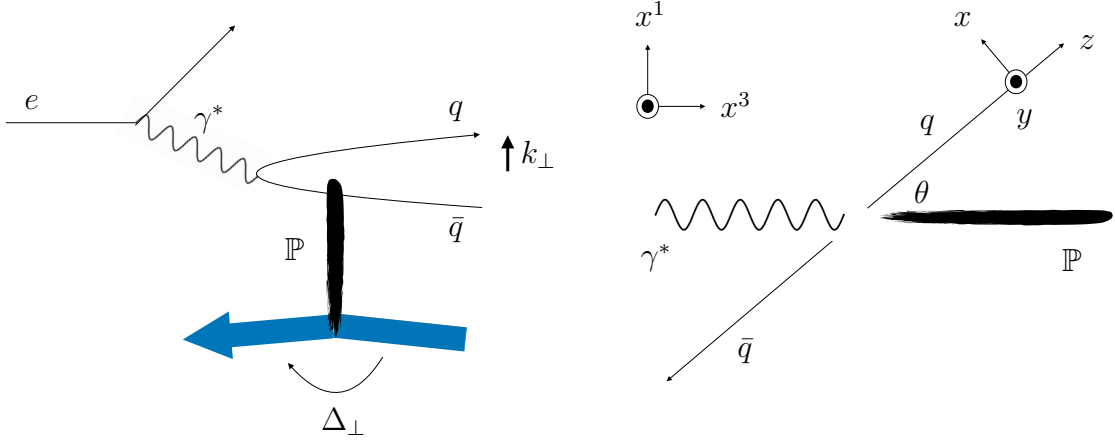
\begin{figure}
\begin{center}
\hspace{-10mm}
\begin{overpic}[width=0.5\textwidth]{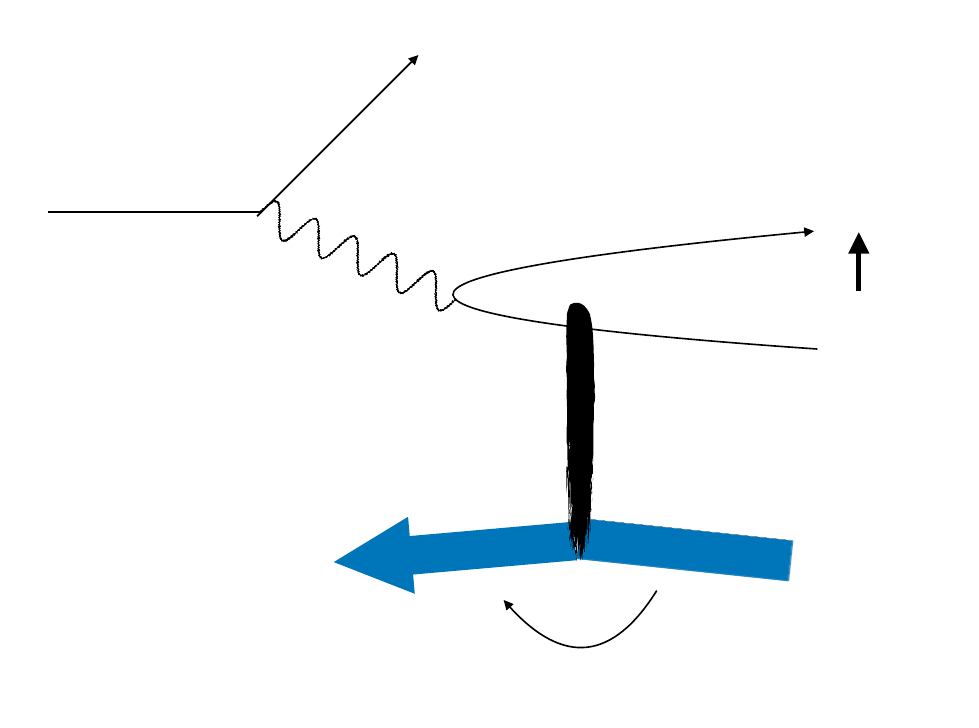}
\put(78,34){\large $\bar{q}$}
\put(78,54){\large $q$}
\put(57,2){\large $\Delta_\perp$}
\put(51,30){\large $\mathbb{P}$}
\put(90,46){\large $k_\perp$}
\put(8,55){\large $e$}
\put(36,52){\large $\gamma^*$}
\end{overpic}
\begin{overpic}[width=0.47\textwidth]{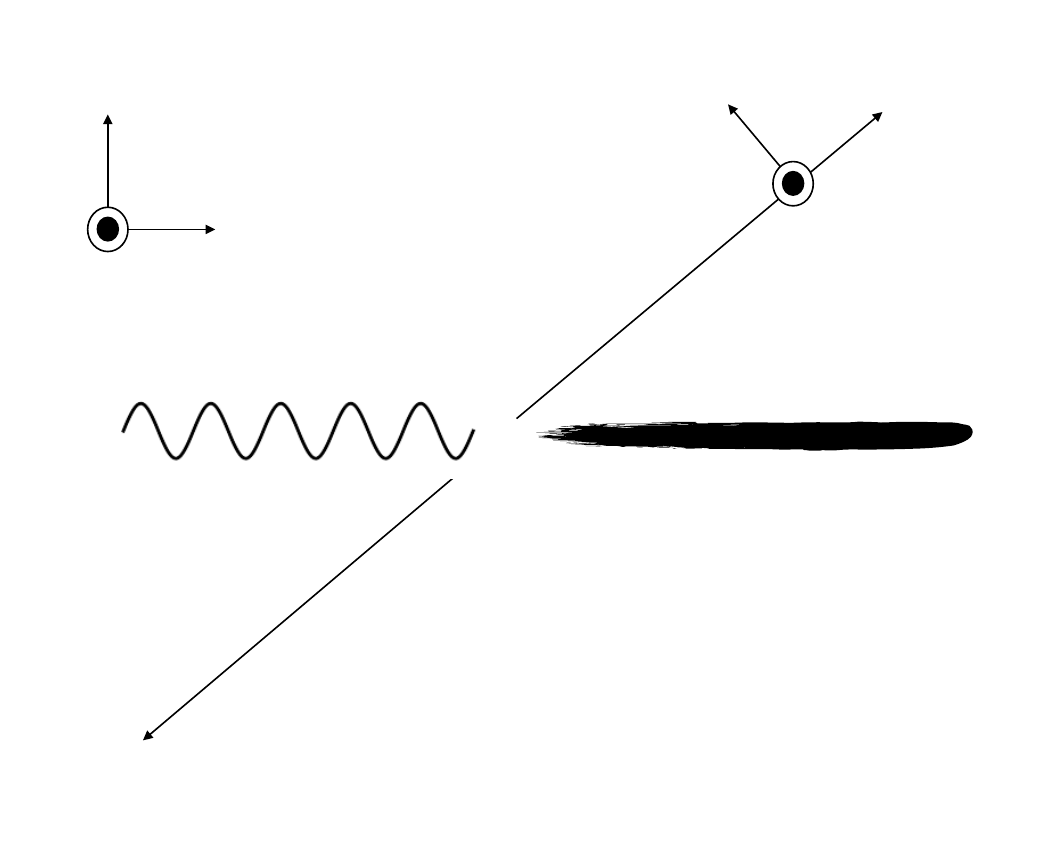}
\put(26,15){\large $\bar{q}$}
\put(58,55){\large $q$}
\put(9,31){\large $\gamma^*$}
\put(23,57){\large $x^3$}
\put(8,73){\large $x^1$}
\put(86,69){\large $z$}
\put(64,73){\large $x$}
\put(75,56){\large $y$}
\put(83,30){\large $\mathbb{P}$}
\put(57,42){\large $\theta$}
\end{overpic}
\caption{Left: Diffractive $q\bar{q}$ production in the dipole frame. Right: $q\bar{q}$ center-of-mass frame }
\label{DA}
\end{center} \end{figure}

For definiteness, in this paper we focus on the diffractive production of $q\bar q$ pairs (see, e.g.,~\cite{Diehl:1994wz,Bartels:1996ne,Nikolaev:2003zf,Braun:2005rg,Altinoluk:2015dpi,Hatta:2016dxp,Boussarie:2019ero,Pang:2026lsr}), for which the spin-density matrix has been computed in~\cite{Fucilla:2025kit,Hatta:2025obw}. The discussion  can be straightforwardly extended to the inclusive case, with the corresponding  spin-density matrix \cite{Qi:2025onf,Fucilla:2026mkg}. The physical picture of diffractive production is depicted in Fig.~\ref{DA}. The electron first emits a virtual photon which then splits into a $q\bar{q}$ pair and   exchanges color singlet objects  in the $t$-channel, such as the two-gluon state, the quark-antiquark state and their higher order  generalizations. We collectively refer to them as `Pomeron' $\mathbb{P}$ and regard the subprocess   $\gamma^*+\mathbb{P}\to q(k)+\bar{q}(k')$ as   $2\to 2$ scattering. We initially work in the so-called `dipole frame'  in which  the virtual photon and the target are collinear and the photon has large momentum  along the $x^3$ axis. The $\gamma^*p$ (or $\gamma^*A$) cross section can be generally written as, for each quark flavor $q$,
\beq
\frac{d\sigma^q_{\rm DIS}}{d\xi d^2k_\perp d^2\Delta_\perp}=\frac{d\sigma^q_T}{d\xi d^2k_\perp d^2\Delta_\perp}+\varepsilon\frac{d\sigma^q_L}{d\xi d^2k_\perp d^2\Delta_\perp} \equiv A^q_T+\varepsilon A^q_L \; ,
\label{omit}
\eeq
where $\varepsilon$ is the ratio of  the longitudinal ($L$) and transverse ($T$) virtual photon fluxes, $\xi$ ($\bar{\xi}=1-\xi$) is the longitudinal momentum fraction of the photon carried by the quark (antiquark), and  $\vec{\Delta}_\perp$ is the recoil momentum  of the target proton/nucleus.
We approximate $\Delta_\perp\approx 0$ since  the cross section is sharply peaked there. It then follows that
$\vec{k}_\perp=-\vec{k}'_\perp$, and the pair has invariant mass
\beq
{\cal M}^2=\frac{k_\perp^2+m^2_q}{\xi\bar{\xi}}.
\eeq

Under these assumptions, the final state closely resembles that of $e^+e^-$ annihilation in a highly boosted frame. To make the discussion parallel,  we fix ${\cal M}$ by inserting  $1=\int d{\cal M}^2\delta({\cal M}^2-\frac{k_\perp^2+m_q^2}{\xi\bar{\xi}})$ and integrate over  $k_\perp$
\beq
\left.\frac{d\sigma_{\rm DIS}^q}{d{\cal M}^2d\xi d^2\Delta_\perp} \right|_{\Delta_\perp=0}&=&\int d^2k_\perp\left.\frac{d\sigma_{\rm DIS}^q}{d\xi d^2k_\perp d^2\Delta_\perp}\right|_{\Delta_\perp=0} \delta \left( {\cal M}^2-\frac{k_\perp^2+m_q^2}{\xi\bar{\xi}} \right) \nn
&=&\pi \xi\bar{\xi}\left.\frac{d\sigma_{\rm DIS}^q}{d\xi d^2k_\perp d^2\Delta_\perp}\right|_{\Delta_\perp=0}  ,
\eeq
where $k^{2}_\perp =\xi\bar{\xi}{\cal M}^2-m^2_q$.  The $d^2k_\perp$ integral eliminates terms in $d\sigma/d^2k_\perp$ that depend on the angle of $\vec{k}_\perp$ with respect to the lepton plane. We may then take $\vec{k}_\perp$ to be along the $x^1$ axis,  boost along the $x^3$ axis to reach the CM frame of the $q\bar{q}$ pair, and finally  rotate around the $x^2$ axis by angle $\theta$ (see Fig.~\ref{DA})
\beq
\cos\theta = \frac{(2\xi-1){\cal M}}{\sqrt{{\cal M}^2-4m^2_q}} =\frac{2\xi-1}{\beta_q}.
\eeq
In this way, we have  arrived at the `$xyz$' frame as in the final state of $e^+e^-$ annihilation.
The analogy becomes complete by switching variables
 from $\xi$ to $\theta$
\beq
\left.\frac{d\sigma^q_{\rm DIS}}{d{\cal M}^2d\cos\theta d^2\Delta_\perp} \right|_{\Delta_\perp=0} =\pi \beta_q \frac{ {\cal M}^2\sin^2\theta +4m_q^2\cos^2\theta}{8{\cal M}^2}\left.\frac{d\sigma^q_{\rm DIS}}{d\xi d^2k_\perp d^2\Delta_\perp}\right|_{\Delta_\perp=0}.
\eeq

Let us discuss the fragmentation of the quark and antiquark in this frame. The angular distribution of hadron pairs  $q\bar{q}\to h\bar{h}X$ is given by
\beq
&& \left.\frac{zz'd\sigma^h_{\rm DIS}}{dzdz' d\phi_Pd\phi_{P'}d{\cal M}^2d\cos\theta d^2\Delta_\perp} \right|_{\Delta_\perp=0}= \sum_q\left.\frac{zz'd\sigma^q_{\rm DIS}}{d{\cal M}^2d\cos\theta d^2\Delta_\perp} \right|_{\Delta_\perp=0}\times \frac{1}{4\pi^2}
 \Biggl\{z D_1^{h/q}(z)z'D^{\bar{h}/\bar{q}}_1(z') \nn && \qquad   -B^i_{\rm DIS}\epsilon^{ij}\hat{P}^j H_1^{\perp(1)}z'\bar{D}_1 +\bar{B}^i_{\rm DIS}\epsilon^{ij}\hat{P}'^j zD_1\bar{H}_1^{\perp(1)}
- \epsilon^{il} \hat{P}^l_\perp \epsilon^{jm} \hat{P}'^m_\perp C_{\rm DIS}^{ij} H_1^{\perp(1)} \bar{H}_1^{\perp(1)} \Biggr\} ,
 \eeq
 where
\beq
B^a_{\rm DIS}= \frac{A_TB^a_T+\varepsilon A_LB^a_L}{A_T+\varepsilon A_L}, \qquad \bar{B}^a_{\rm DIS}= \frac{A_T\bar{B}^a_T+\varepsilon A_L\bar{B}^a_L}{A_T+\varepsilon A_L}, \qquad C^{ab}_{\rm DIS}=\frac{A_TC^{ab}_T+\varepsilon A_LC^{ab}_L}{A_T+\varepsilon A_L}. \label{disbb}
\eeq
The unpolarized cross sections $A_{L/T}$ are as defined in (\ref{omit}).
The cross section for dihadron pair production is similarly given by (\ref{most}) with trivial modifications, but this most general form is not needed. In practice, one can conduct a targeted search  for each matrix element
\beq
&& \frac{zz' d\sigma^h_{\rm DIS}}{ dz dz'd\phi_{RP} d\phi'_{RP} d{\cal M}^2d\cos\theta d^2\Delta_\perp} =\sum_q\left.\frac{zz'd\sigma^q_{\rm DIS}}{d{\cal M}^2d\cos\theta d^2\Delta_\perp} \right|_{\Delta_\perp=0} \nn
&& \qquad \times \frac{1}{(2\pi)^2}\Biggl\{zD_1^{(0)}(z)z'\bar{D}_1^{(0)}(z')+C_{\rm DIS}^{zz}G_1^{\perp(1)}(z)\bar{G}_1^{\perp(1)}(z')  \sin(\phi_{RP})\sin(\phi'_{RP}) \Biggr\},
\eeq
\beq
&& \frac{zz'd\sigma^h_{\rm DIS}}{dz dz'd\phi_R d\phi'_{RP}d{\cal M}^2d\cos\theta d^2\Delta_\perp}=\sum_q\left.\frac{zz'd\sigma^q_{\rm DIS}}{d{\cal M}^2d\cos\theta d^2\Delta_\perp} \right|_{\Delta_\perp=0} \\
&& \qquad  \times \frac{1}{(2\pi)^2} \left\{zD_1^{(0)}(z) z'\bar{D}^{(0)}_1(z')+\bar{B}^z zD_1^{(0)}(z)\sin\phi'_{RP}\bar{G}_1^{\perp(1)}(z')- C_{\rm DIS}^{iz}\epsilon^{ij}\hat{R}^j_\perp H_1^{\sphericalangle(1)}(z) \sin \phi'_{RP}\bar{G}_1^{\perp (1)}(z')\right\}, \notag
\eeq
and similarly for $C^{zi}_{\rm DIS}$ and $B^z$. Note that, in practice, $B^z=\bar{B}^z=0$ and
\beq
&& B^i\epsilon^{ij}\hat{P}^j\to -B^y \cos\phi_P,\nn
&& C^{ij} \epsilon^{il} \hat{P}^l_\perp \epsilon^{jm} \hat{P}'^m_\perp \to C^{xx}\sin\phi_P \sin\phi_{P'} +C^{yy}\cos\phi_P\cos\phi_{P'},
\nn
&& C^{iz}\epsilon^{ij}\hat{R}^j_\perp \to C^{xz} \sin \phi_R,
\eeq
due to parity. But this should be experimentally confirmed rather than assumed in quantum tomography.

In diffractive production, the spin-averaged cross sections $A_{L/T}$ have been known for a long time \cite{Bartels:1996ne,Nikolaev:2003zf,Braun:2005rg}, but  the spin density matrix elements $B_{L/T}^a,\bar{B}_{L/T}^a,C_{L/T}^{ab}$  have been  computed only recently in \cite{Fucilla:2025kit} in the $k_T$ factorization framework, and in \cite{Hatta:2025obw} in the collinear factorization framework. For definiteness, here we use $k_T$ factorization  and discuss the implications of the results obtained in   \cite{Fucilla:2025kit}, relegating the explicit  formulas to Appendix \ref{appB}.
First, $B^a=\bar{B}^a=0$ in the high energy (`eikonal') approximation, and this will be assumed in the following. We however note that  the normal components  $B^y$ and $\bar{B}^y$ need not vanish in  QCD in general, and indeed, they  are nonvanishing at the sub-eikonal level \cite{Hatta:2026dqs}.

Second, in the massless limit, the $C$-matrices  simplify  dramatically
\beq
C_L=\begin{pmatrix} -1 & 0 & 0 \\ 0 & 1 & 0 \\ 0 & 0 & 1\end{pmatrix}, \qquad  C_T= \begin{pmatrix} \frac{2\xi\bar{\xi}}{\xi^2+\bar{\xi}^2} & 0 & 0 \\ 0 & -\frac{2\xi\bar{\xi}}{\xi^2+\bar{\xi}^2} & 0 \\
0 & 0 & 1
\end{pmatrix}. \label{m0}
\eeq
Taken individually, $C_L$ denotes the Bell state
\beq
|\Phi^-\rangle=\frac{|++\rangle-|--\rangle}{\sqrt{2}},
\label{bell1}
\eeq
where $\pm$ denotes the spin projection along the $z$-axis.
In fact, $C_T$  is identical to the massless limit of (\ref{xyz}), and reduces to another Bell state at the symmetric point $\xi=\bar{\xi}=\frac{1}{2}$
\beq
|\Phi^+\rangle =\frac{|++\rangle+|--\rangle}{\sqrt{2}} \; . \label{bell2}
\eeq
The linear combination $C_{\rm DIS}$ in general represents a mixed state. Depending on the strength of $\varepsilon A_L$ ($\varepsilon \approx 1$ in typical EIC kinematics), the angular distribution   can be significantly modified from its  canonical form $\cos(\phi+\phi')$. In particular, one can always find a locus in phase space where
\beq
\frac{2\xi\bar{\xi}}{\xi^2+\bar{\xi}^2} A_T=\varepsilon A_L \; . \label{zero}
\eeq
Along this locus, $C_{\rm DIS}\approx {\rm diag} (0,0,1)$, meaning the complete disappearance of the Collins effect, or any angular correlations from the Collins FF and $H_1^\sphericalangle$. Only the dihadron correlation $G_1^{\perp}\bar{G}_1^\perp \sin \phi_{RP}\sin\phi'_{RP}$ survives with the maximal coefficient  $C^{zz}_{\rm DIS}=1$.

Finally, our main interest is massive $q\bar{q}$ production.   In this case, the $C_T$-matrix depends on the nonperturbative `dipole T-matrix' which incorporates the gluon saturation effect of the target nucleus. Nevertheless, independently of models, the matrix elements satisfy the following relations  \cite{Fucilla:2025kit}\footnote{The equalities hold in the high energy (eikonal) approximation. At finite energy, there are calculable  corrections to these relations \cite{Hatta:2026dqs}.}
\beq
(C^{xx}_{T})^2+(C^{xz}_{T})^2+(C^{zx}_{T})^2+(C^{zz}_{T})^2-(C^{yy}_{T})^2=1 \; , \label{re1}
\eeq
\beq
C^{yy}_{T}+C^{xx}_{T}C^{zz}_{T}-C^{xz}_{T}C^{zx}_{T}=0 \; . \label{re2}
\eeq
Remarkably, when these relations hold,  entanglement \cite{Peres:1996dw,Horodecki:1997vt}  and Bell-nonlocality \cite{Horodecki:1995nsk} become necessary and sufficient to each other, although usually the latter is a subset of the former. The criterion of entanglement ($=$ Bell-nonlocality) is then simply
\beq
C^{yy}_T<0 \; .
\eeq
This can be experimentally verified in photo-production ($A_L=0$). In fact, one can even  test the relations (\ref{re1}) and (\ref{re2}) by measuring  all components of $C_T$.

Moreover, if there is a point in phase space where  $C_T^{yy}=-1$, then from (\ref{re1}) and (\ref{re2}),
\beq
(C^{xx}_{T})^2+(C^{xz}_{T})^2+(C^{zx}_{T})^2+(C^{zz}_{T})^2=2, \quad
C^{xx}_{T}C^{zz}_{T}-C^{xz}_{T}C^{zx}_{T}=1,
\eeq
so that
\beq
C_T=\begin{pmatrix} \cos\psi & 0 & -\sin \psi \\ 0 & -1 & 0 \\ \sin\psi & 0 & \cos\psi\end{pmatrix}, \qquad {\rm det}C_T=-1. \label{psi}
\eeq
This represents a maximally entangled state. An immediate prediction is that near this kinematical point, the Collins asymmetry is replaced by
\beq
\cos(\phi_P+\phi_{P'}) \to \cos\phi_P \cos\phi_{P'} -\cos\psi \sin\phi_P\sin \phi_{P'}.  \label{dec}
\eeq
The value of $\psi$ depends on the structure of the target.
As shown in \cite{Fucilla:2025kit} (see also Appendix~\ref{appB}), there exists an important class of models where the maximal value   $C_T^{yy}=-1$ is attained
at  $\xi=\frac{1}{2}$ ($\cos\theta=0$) and
\beq
k^2_\perp \approx \frac{Q^2}{4}+m_q^2 \; \to Q\approx \sqrt{{\cal M}^2-8m_q^2}, \label{rem}
\eeq
at which the $C_T$-matrix takes the form
\beq
C_T^{ab}\approx \begin{pmatrix} 0 & 0 & -1\\ 0 & -1 & 0 \\ 1 & 0 & 0\end{pmatrix},
\label{max}
\eeq
where the approximate relation becomes exact in the heavy quark mass limit $m_q\to \infty$. Therefore, $\psi\approx \frac{\pi}{2}$ in this case and the Collins asymmetry is  modified as
\beq
\cos(\phi_P+\phi_{P'})H_1^\perp \bar{H}_1^\perp \to \cos\phi_P \cos\phi_{P'} H_1^\perp \bar{H}_1^\perp. \label{dec0}
\eeq
At the same time, the off-diagonal correlations
\beq
\sin\phi_R \sin\phi'_{RP} H_1^\sphericalangle \bar{G}_1^\perp , \qquad \sin\phi_{RP} \sin\phi_{R'}G_1^\perp \bar{H}_1^\sphericalangle,
\eeq
generate maximal signals, whereas the   $G_1^\perp\bar{G}_1^\perp$ correlation vanishes. It is intriguing to notice that, somewhat counterintuitively, this novel form of  maximal entanglement  leads to the disentanglement of $\phi_P$ and $\phi_{P'}$ in the Collins sector (\ref{dec}).

\section{Numerical results}

\subsection{Preliminary}

In this section, we present numerical results on the size of asymmetries discussed in the previous sections. The input fragmentation functions are as follows:

\noindent (i) The single-hadron unpolarized FFs  for
light quarks $q=u,d,s$ into charged pions $D_1^{\pi^\pm/q}(z)=D_1^{\pi^\mp/\bar{q}}(z)$  are taken from the DSS
leading-order parameterization \cite{deFlorian:2007aj}.

\noindent (ii) The Collins FFs for light quarks into charged pions $H_1^{\perp \pi^\pm/q}(z,P_{\pi \perp}^2)=H_1^{\perp \pi^\mp/\bar{q}}(z,P_{\pi \perp}^2)$ are taken from the \texttt{JAM3D-22} extraction \cite{Gamberg:2022kdb}; see also~\cite{Anselmino:2015sxa,Kang:2015msa,Zeng:2023nnb}. The moment (\ref{half}) is computed with the fitted Gaussian widths and $M_\pi=0.134$ GeV from the \texttt{JAM3D-22} extraction.

\noindent (iii) At the EIC, `massive quark' in practice means the charm quark. We take the charge-inclusive   unpolarized FFs $D_1^{D/c}(z)\equiv D_1^{D^+/c}(z)+D_1^{D^-/c}(z)=D_1^{D/\bar{c}}(z)$ from the KKKS08 OPAL-only NLO general-mass set \cite{Kneesch:2007ey}; see also~\cite{Salajegheh:2019nea}.

To our knowledge, there has been no extraction of the Collins FFs for the charm quark. We therefore use the following model
\beq
H_1^{\perp D/c(1)}(z) =  \frac{D_1^{D/c}(z)}{D_1^{\pi^+/u}(z)}H_1^{\perp \pi^+/u(1)}(z). \label{model}
\eeq

\noindent (iv)
While global fits to the interference FF
$H_1^{\sphericalangle}$ are  available \cite{Courtoy:2012ry,Cocuzza:2023vqs,Cocuzza:2023oam} and they can be used to access the $C^{ij}$ components, the same can be achieved via the Collins FFs as we have seen. The other components of the $C$-matrix require the $G_1^\perp$ dihadron FFs. Experimental evidence for a nonzero $G_1^\perp$ has been reported by the CLAS12 Collaboration \cite{Hayward:2021psm}, although its quantitative extraction  is not yet available. We therefore do not perform numerical calculations with dihadron FFs in this paper.

\begin{figure}[thb]
\centering
\includegraphics[width=0.5\textwidth]{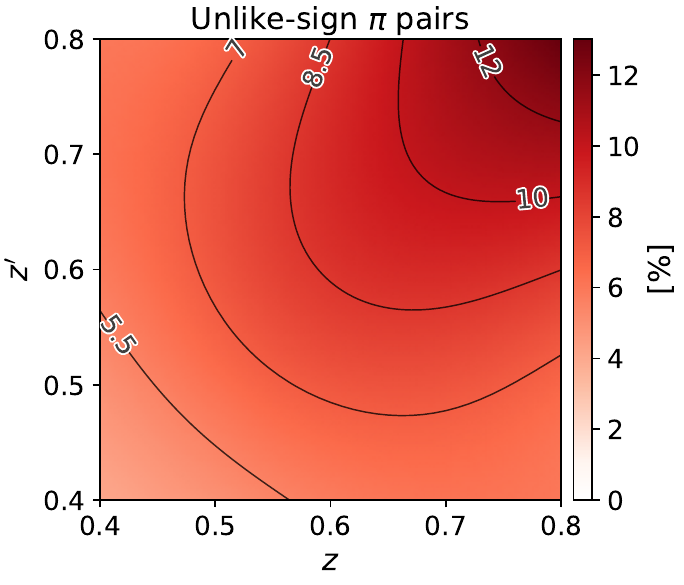}
\includegraphics[width=0.49\textwidth]{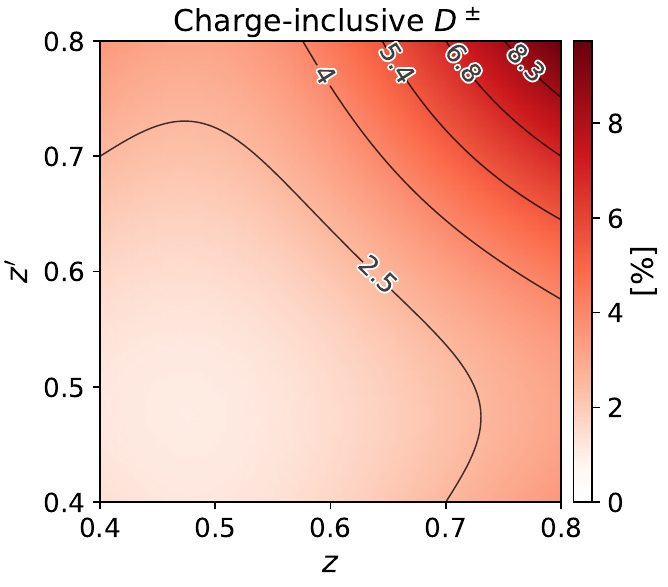}

\caption{ Collins asymmetry coefficients $A_{12}(z,z')$ at $\mu_F=10.58$~GeV. The left and right panels show the light-quark asymmetries for unlike-sign charged-pion pairs and the charm asymmetries inclusive of $D^{\pm}$.  Colors represent the mean value average over replicas. The uncertainties relative to the mean defined with the half-width of the $68\%$-confidence interval is about $12$--$27\%$ for pions and $26$--$55\%$ for D-mesons, which are estimated with  replicas of the fragmentation functions (not shown in the plot).} \label{fig:unintegrated-collins-2d}
\end{figure}

Let us quickly reproduce the Collins asymmetry in $e^+e^-$ annihilation at Belle/BaBar energy $\sqrt{s}=10.58$ GeV. Since the FFs have been partly constrained by the Belle/BaBar data, the only new discussion here is the Collins effect in D-mesons (in our model), which is yet to be measured experimentally.
We consider unlike-sign pairs $\pi^+\pi^-$ for pions and charge-inclusive pairs $(D^++D^-)(D^++D^-)$  for D-mesons,\footnote{Ideally, one should detect opposite charge pairs $D^+D^-$ to  better discriminate between quark and antiquark jets, which is important for the measurement of off-diagonal components, see (\ref{off}) for example. However, only the charge-inclusive $c\to D^++D^-$ FFs are available to us at the moment. One can also consider the channel $c\to D^0$ and $\bar{c}\to\bar{D}^0$ which enjoys a larger fragmentation probability. }
  and parametrize the angular distribution in the  form
\beq
\frac{d\sigma^{\pi/D}}{dzdz'd\phi_P d\phi_{P'}d\cos\theta}\propto 1+A_{12}^{\pi/D}(z,z') \frac{\sin^2\theta}{1+\cos^2\theta+\frac{4m_q^2}{s}}\left(\cos(\phi_{P}+\phi_{P'})-\frac{4m^2_q}{s}\cos(\phi_P-\phi_{P'})\right), \label{eepara}
\eeq
\beq
&& A_{12}^{\pi}(z,z')=
\frac{\sum_q e_q^2(H_1^{\perp \pi^+/q(1)}(z)H_1^{\perp \pi^-/\bar{q}(1)}(z')+(q\leftrightarrow \bar{q}))}
{\sum_qe_q^2 (z D_1^{\pi^+ /q}(z) z'D_1^{\pi^- / \bar q}(z')+(q\leftrightarrow \bar{q}))}, \nn
&&  A_{12}^{D}(z,z')=
\frac{H_1^{\perp D/c(1)}(z)H_1^{\perp D/\bar{c}(1)}(z')}
{ zD_1^{D/c}(z)z'D_1^{D/\bar{c}}(z')},
\eeq
where $m_q=0$ for $\pi$ and $m_q=m_c$ for $D$.
We neglect the contributions $u,d,s\to D$ and $c\to \pi$.
Actually, in the present model (\ref{model}), the ratio $H_1^{\perp(1)}/D_1$ is the same for $c\to D$ and $u\to \pi^+$. Thus the only difference between $A_{12}^\pi$ and $A_{12}^D$ comes from the combinatorics of `favored' (such as $u\to \pi^+$) and `disfavored' (such as $u\to \pi^-$)  contributions.
The result is shown in Fig.~\ref{fig:unintegrated-collins-2d}. The leading asymmetry $\cos(\phi_P+\phi_{P'})$ is sizable, reaching 10\% in the highest $z,z'$ bins. However, the subleading $\cos(\phi_P-\phi_{P'})$ modulation is challenging to measure due to the suppression factor    $4m_c^2/s=0.0804$.

\subsection{Azimuthal  asymmetries in DIS}

We now discuss the Collins effect in DIS. As we have seen, the angular dependence on $\theta$, $\phi_P$ and $\phi_{P'}$ can be completely different from the canonical form (\ref{eepara}). We thus consider the more general  parametrization
\beq
d\sigma^{\pi/D}_{\rm DIS}&\propto& 1-A_{12}^{\pi/D}(z,z')\left(C^{xx}_{\rm DIS}\sin\phi_P\sin \phi_{P'}+C^{yy}_{\rm DIS}\cos\phi_P\cos\phi_{P'}\right) \nn
&=&1+A_{12}^{\pi/D}(z,z')\left(\frac{C^{xx}_{\rm DIS}-C^{yy}_{\rm DIS}}{2}\cos(\phi_P+\phi_{P'})-\frac{C^{xx}_{\rm DIS}+C^{yy}_{\rm DIS}}{2}\cos(\phi_P-\phi_{P'})\right) ,
\eeq
in terms of the spin density matrix elements. In $e^+e^-$ annihilation, $C^{xx}\approx -C^{yy}$. In DIS, $C^{xx}$ and $C^{yy}$ are complicated functions of $Q^2$, $\varepsilon$, $m_q$, $\theta$ and ${\cal M}$ (alternatively, $\theta,{\cal M}\to \xi,k_\perp$).  However, the dependence on $z,z'$ factorizes and is the same as in $e^+e^-$ annihilation (Fig.~\ref{fig:unintegrated-collins-2d}). We thus focus on the behavior of $C^{xx}_{\rm DIS}$ and $C^{yy}_{\rm DIS}$.

In Appendix  \ref{appB}, we have collected necessary formulas to compute the matrix $C^{ab}_{\rm DIS}$. They require the `dipole T-matrix' as a model input.  To study the dependence on different targets and different models, we employ  the Golec-Biernat-Wusthoff (GBW) model for the proton \cite{Golec-Biernat:1998zce}
and the McLerran-Venugopalan (MV) model for a heavy nucleus  \cite{McLerran:1993ni}. They are given in (\ref{gbw}) and (\ref{mv}), respectively.

\begin{figure}[h]
\centering
\includegraphics[width=\textwidth]{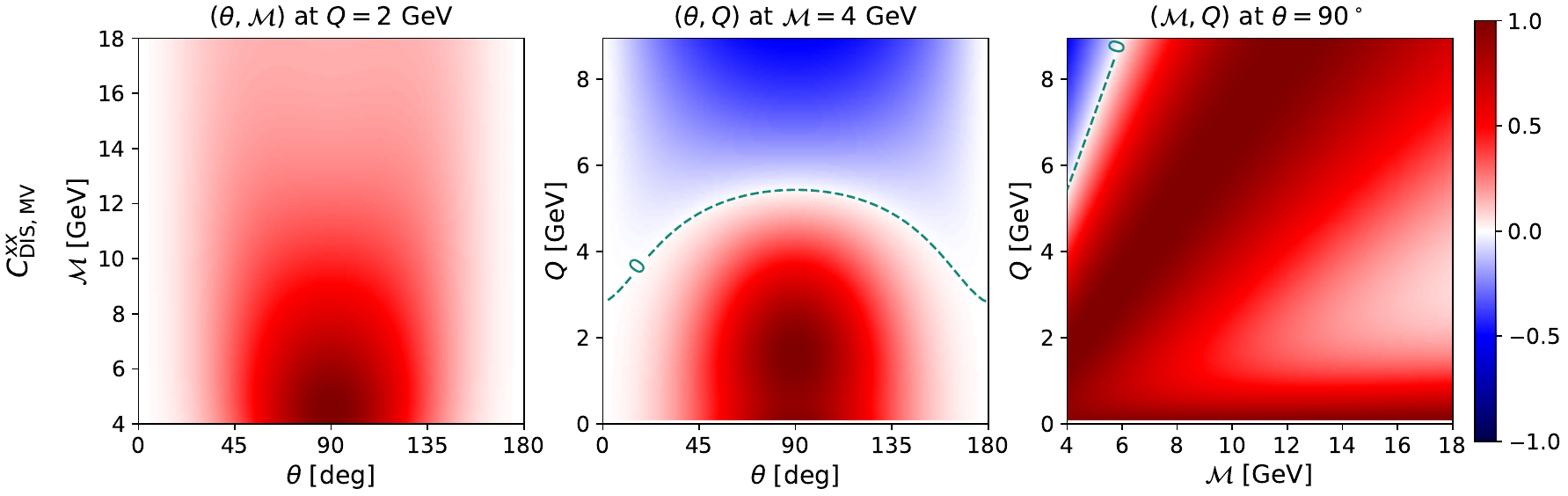}
\caption{The spin correlation $C^{xx}_{\rm{DIS}}=-C^{yy}_{\rm{DIS}}$ for massless quarks with $\varepsilon=1$ and the MV gold-nucleus dipole at $Q_{sA}^2=1.5~\mathrm{GeV}^2$. From left to right, the planes are $(\theta,{\cal M})$ at $Q=2$~GeV, $(\theta,Q)$ at ${\cal M}=4$~GeV, and $({\cal M},Q)$ at $\theta=90^\circ$. Dashed lines are the zero contours where the correlation and corresponding Collins asymmetry vanish.} \label{fig:dis-light-spin}
\end{figure}

First consider $\pi^+\pi^-$ production from massless quarks $m_q=0$. In this case, $C^{xx}_{\rm DIS}=-C^{yy}_{\rm DIS}$ exactly, but $C^{xx}_{\rm DIS}$ can  significantly differ from the familiar function $\frac{\sin^2 \theta}{1+\cos^2\theta}$ due to the mixing of the longitudinally polarized virtual photon.    In Fig.~\ref{fig:dis-light-spin}, we plot $C^{xx}_{\rm DIS}$ obtained in the MV model in the $(\theta,{\cal M})$ (left), $(\theta,Q)$ (middle) and $({\cal M},Q)$ (right) planes at fixed $\varepsilon=1$. In the blue region, $C^{xx}_{\rm DIS}$ is negative because the longitudinal photon contribution dominates over the transverse one. The teal curves denote the boundary   $C^{xx}_{\rm DIS}=0$ along which the condition (\ref{zero}) is met and the dependence on $\phi_P,\phi_{P'}$ disappears. In diffractive production, on general grounds,  one expects that the longitudinal contribution dominates over the transverse one in the large-$Q$ region, and this is most clearly seen in the middle plot. However, the actual situation is more complicated because there are two competing scales $Q$ and ${\cal M}$.  Maximally entangled Bell states (\ref{bell1}) and (\ref{bell2}) are realized in the darkest blue ($C^{xx}_{\rm DIS}=-1$) and darkest red ($C^{xx}_{\rm DIS}=1$) regions, where the strength of the $\cos(\phi_P+\phi_{P'})$ asymmetry becomes maximal.

\begin{figure}[th]
\centering
\includegraphics[width=\textwidth]{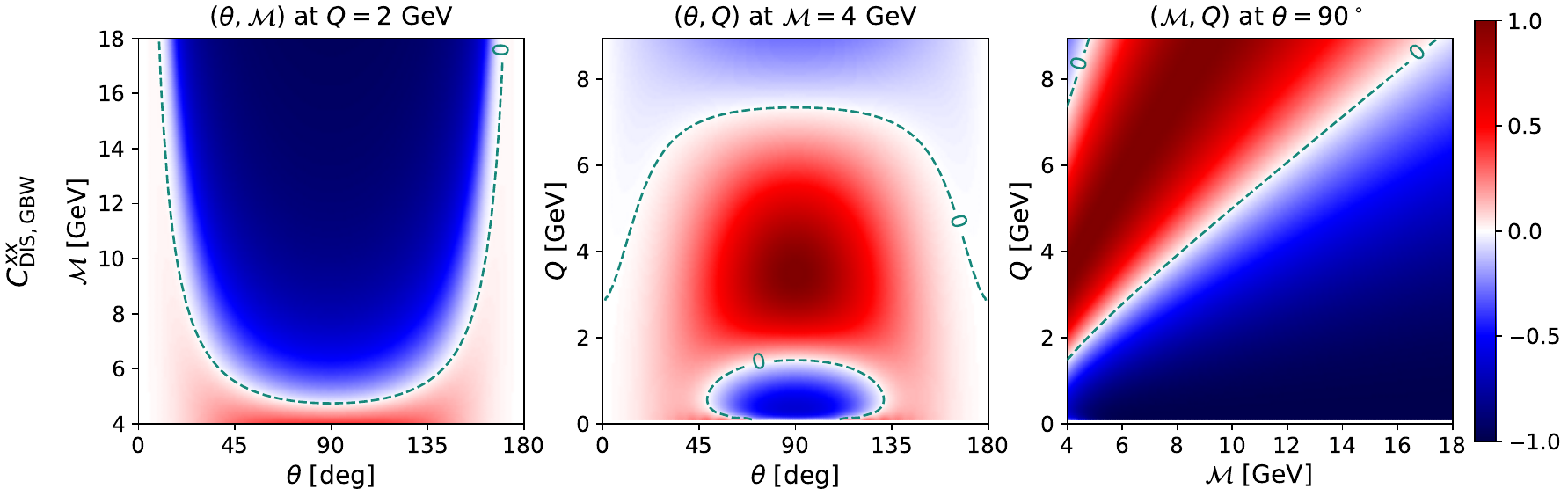}
\caption{The same as Fig.~\ref{fig:dis-light-spin}, but in the GBW model.
} \label{lightGBW}
\end{figure}

The corresponding plots for the GBW model in Fig.~\ref{lightGBW} show a qualitatively different behavior. Although the longitudinal contribution vanishes at $Q=0$, somewhat surprisingly,  it immediately dominates over the transverse one as soon as $Q>0$.\footnote{This can be understood by noticing that the integrands in  (\ref{t1t2}) are singular at $Q=m=0$. The sum $T_1+T_2$ in (\ref{at}) is finite after a large cancellation.}   The transverse contribution dominates for  intermediate values of $Q$  until it is again taken over by the longitudinal contribution in the large-$Q$ region. Thus, there are two curves of $C^{xx}_{\rm DIS}=0$ in this model, meaning that the sign of the $\cos(\phi_P+\phi_{P'})$ asymmetry changes twice.  Future measurements  should be able to  confirm or exclude this peculiar behavior.

\begin{figure}[th]
\centering
\includegraphics[width=\textwidth]{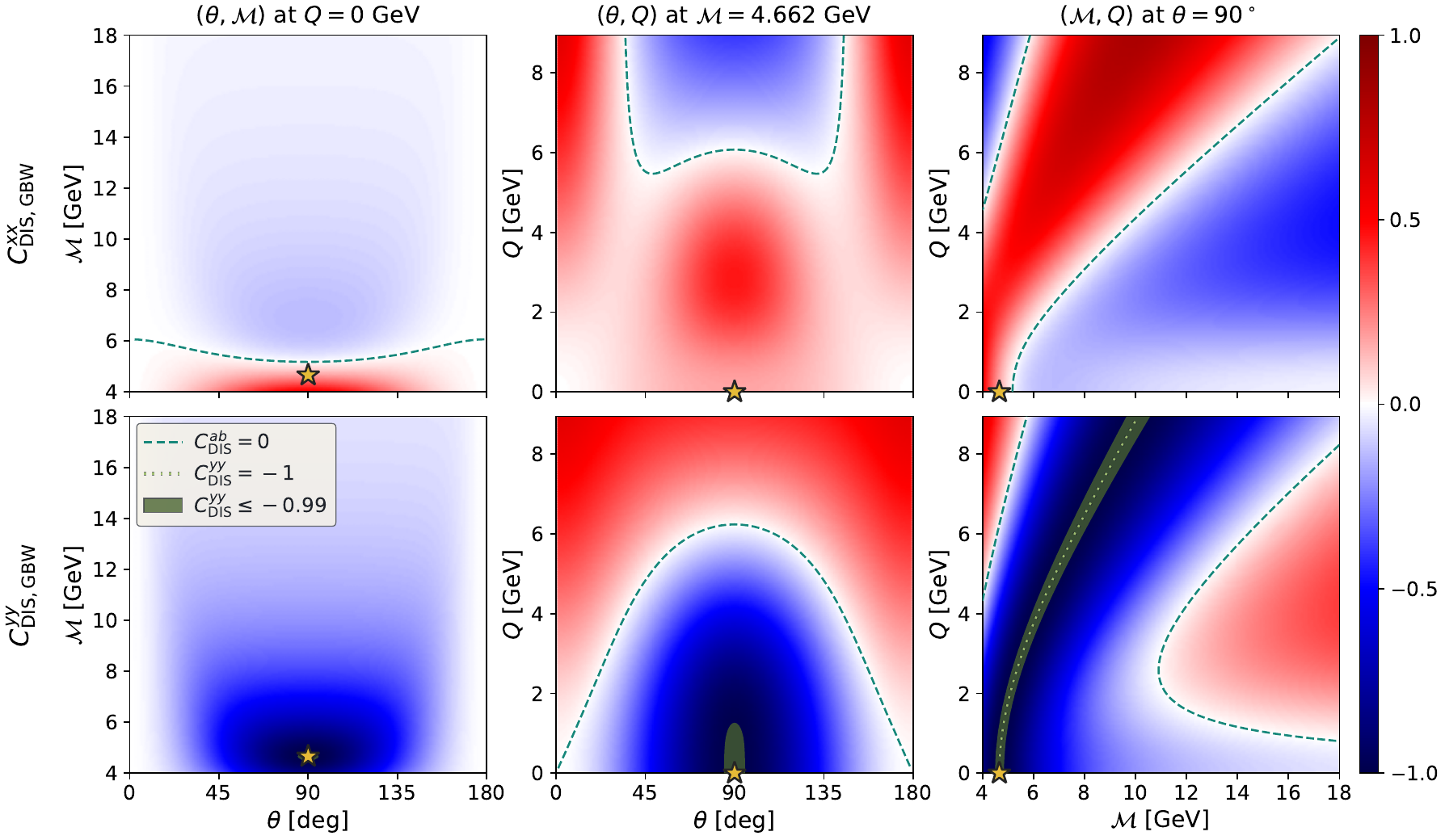}
\caption{The  spin correlation $C^{xx}_{\rm{DIS}}$ and $C^{yy}_{\rm{DIS}}$ for charm quark in diffractive DIS with $m_c=1.5$ GeV, $\varepsilon=1$, and the GBW proton dipole at $Q_s^2=1~\mathrm{GeV}^2$. From left to right, the planes are $(\theta,{\cal M})$ at $Q=0$~GeV, $(\theta,Q)$ at ${\cal M}=4.662$~GeV, and $({\cal M},Q)$ at $\theta=90^\circ$. Dashed lines are the zero contours where the correlation and corresponding Collins asymmetry vanish. The stars mark projections of the common point where $C^{yy}_{\rm{DIS}}=-1$ and $C^{xx}_{\rm{DIS}}\approx0.172$ at $Q=0$ GeV. Shaded bands represent the near-minimal region $C^{yy}_{\rm DIS}\leq-0.99$.}
\label{fig:dis-charm-GBW}
\end{figure}

\begin{figure}[th]
\centering
\includegraphics[width=\textwidth]{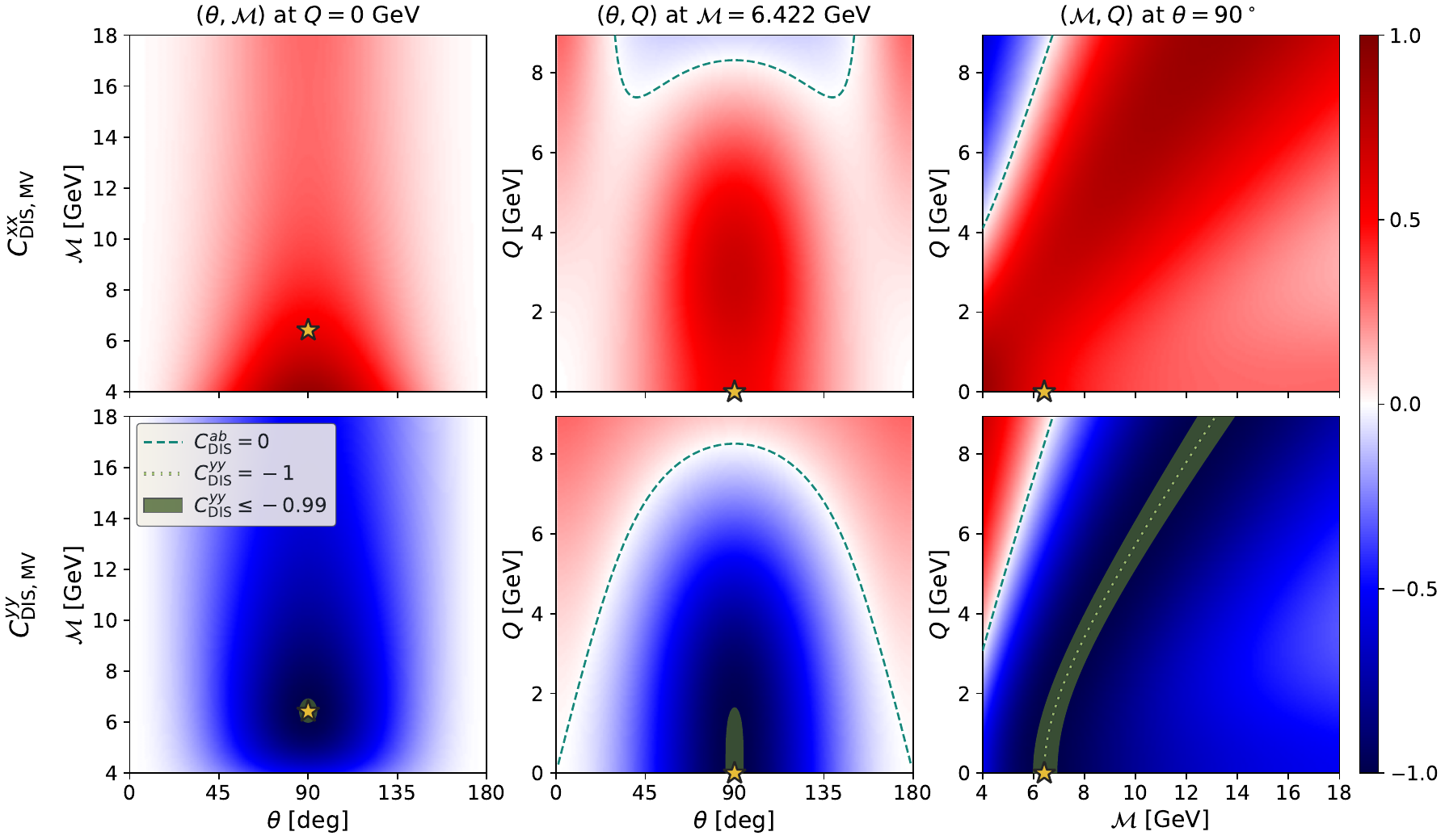}
\caption{The spin correlation $C^{xx}_{\rm{DIS}}$ and $C^{yy}_{\rm{DIS}}$ for charm quark in diffractive DIS with $m_c=1.5$ GeV, $\varepsilon=1$, and the MV gold-nucleus dipole at $Q_{sA}^2=1.5~\mathrm{GeV}^2$. From left to right, the planes are $(\theta,{\cal M})$ at $Q=0$~GeV, $(\theta,Q)$ at ${\cal M}=6.422$~GeV, and $({\cal M},Q)$ at $\theta=90^\circ$. Dashed lines are the zero contours where the correlation and corresponding Collins asymmetry vanish. The stars mark projections of the common point where $C^{yy}_{\rm{DIS}}=-1$ and $C^{xx}_{\rm{DIS}}\approx0.564$ at $Q=0$ GeV. Shaded bands represent the near-minimal region $C^{yy}_{\rm DIS}\leq-0.99$. }
\label{fig:dis-charm-mv}
\end{figure}

Next we consider $D$-meson pair production from $c\bar{c}$ pairs. In Fig.~\ref{fig:dis-charm-GBW}, we plot $C^{xx}_{\rm DIS}$ (upper panels) and $C^{yy}_{\rm DIS}$ (lower panels) in the GBW model, and we do the same in Fig.~\ref{fig:dis-charm-mv}  in the MV model.   The left plots are now evaluated in the photoproduction limit $Q=0$. We notice that  $C^{xx}_{\rm DIS}\neq -C^{yy}_{\rm DIS}$ in general, although their signs are roughly anti-correlated. We also see a  rather strong sensitivity to the choice of models. In the GBW model (Fig.~\ref{fig:dis-charm-GBW}), the star at $Q=0$ indicates the maximally entangled state (\ref{max}) characterized by $C^{yy}_{\rm DIS}=-1$.  $C^{xx}_{\rm DIS}\approx 0.172$ is slightly nonzero  because of the finite ($m_q<\infty$) quark mass. Even for $Q>0$, as long as the longitudinal photon contribution is not significant, we recognize the remnant of this state approximately along the curve $Q=\sqrt{{\cal M}^2-8m_q^2}$ (see (\ref{rem})). For small values of $Q$, this  roughly coincides with the curve  $C^{xx}_{\rm DIS}=0$ where the exotic angular dependence $\cos\phi_P\cos\phi_{P'}$ (\ref{dec0}) is seen.
In other parts of phase space, different linear combinations of $\cos(\phi_P+\phi_{P'})$ and $\cos(\phi_P-\phi_{P'})$ are realized.  We find curves along which $C^{yy}_{\rm DIS}=0$, where  $\frac{d\sigma}{d\phi_Pd\phi_{P'}}\sim \sin\phi_P\sin\phi_{P'}$. In the middle plots, in the so-called aligned jet configuration $\theta\sim 0,\pi$ (or $\xi\sim 0,1$), we even find regions where $C^{xx}_{\rm DIS}=C^{yy}_{\rm DIS}$, or  $\frac{d\sigma}{d\phi_Pd\phi_{P'}}\sim \cos(\phi_P-\phi_{P'})$, in striking contrast to the dominance of $\cos(\phi_P+\phi_{P'})$  in $e^+e^-$ annihilation.

Turning to  the MV model (Fig.~\ref{fig:dis-charm-mv}), we again find a maximally entangled point   $C^{yy}_{\rm DIS}=-1$ indicated by a star at a somewhat larger value of ${\cal M}$. However, the $C$-matrix at this point takes the more general form  (\ref{psi})
\beq
C_T=\begin{pmatrix}
0.5636&0&-0.8260\\
0&-1&0\\
0.8260&0&0.5636
\end{pmatrix},
\eeq
instead of the structure   (\ref{max}) identified in \cite{Fucilla:2025kit}.
In Appendix~\ref{appB}, we explain why the argument of  \cite{Fucilla:2025kit} needs to be revised for the MV model. We note, however, that the  state (\ref{max}) was found in \cite{Hatta:2026dqs} in the collinear factorization framework for the proton target. This suggests that the gluon saturation effects in a large nucleus is responsible for the transition from (\ref{max}) to the more general form (\ref{psi}) with $\psi={\cal O}(1)$.  Since  $C^{xx}_{\rm DIS}$ remains sizable along the curve $C^{yy}_{\rm DIS}=-1$,  the deviation from the $\cos(\phi_P+\phi_{P'})$ asymmetry is significant, but not as dramatic as in the previous case.
The present discussion clearly demonstrates that the spin density matrix elements  and the resulting angular correlations are very sensitive to the structure of the target, emphasizing their relevance to 3D tomography.

Note that the region of interest ${\cal M}\sim 4.5-6.5$ GeV for charm production  would corresponds to the same range of the CM energy $\sqrt{s}$ in $e^+e^-$ annihilation. In this low energy region,  final states typically do not look like clean two-jet events. Nevertheless, the BESIII collaboration measured the Collins asymmetry for $\pi^+\pi^-$ pairs at $\sqrt{s}=3.65$ GeV
\cite{BESIII:2015fyw},  working in the asymmetric frame which avoids the determination of the thrust axis. A similar change of frames may be necessary for probing the maximally entangled state in the $c\bar{c}$ sector. The formula (\ref{mod}) can then be used with the corresponding  $C_{\rm DIS}$ matrix. We however anticipate  that the $\phi_P$-independent part, proportional to $C^{xx}+C^{yy}$, might be hard to distinguish from the unpolarized $D_1D_1$ term.   This problem is absent for bottom quark pairs $b\bar{b}$, although  measuring their diffactive production at the EIC is likely very challenging.

Finally, let us roughly estimate the magnitude of the cross section (see also \cite{Braun:2005rg,Mantysaari:2019csc}). Using the GBW model (\ref{gbw}) and assuming the exponential $\Delta_\perp$-dependence $d\sigma/d^2\Delta_\perp \propto e^{-b|\Delta_\perp|^2}$ with  $b=5$ GeV$^{-2}$ \cite{Braun:2005rg}, we find $\sigma(\gamma^* p\to c\bar c p)\sim200$ nb. The most interesting  kinematic window $4.5<{\cal M}<4.8$ GeV and $|\theta-90^\circ|<5^\circ$, where $C^{yy}_{\rm DIS}$ ranges from $-1$ to about $-0.982$ (see Fig.~\ref{fig:dis-charm-GBW}), accounts for roughly $2\times10^{-3}$ of the  total cross section, yielding $\sigma_{\gamma^* p}^{\rm window}\simeq0.4$ nb. For a representative EIC setup with 18 GeV electrons on 275 GeV protons, the equivalent-photon flux (see, e.g., \cite{H1:2020lzc}) integrated over $Q^2<1$ GeV$^2$, $\sqrt{s_{\gamma p}}>20$ GeV, and $y<0.95$ is $\mathcal{O}(0.1)$, yielding $\sigma_{ep}^{\rm window}\sim40$ pb from this narrow kinematic window. With  100 fb$^{-1}$ luminosity, this amounts to $4\times 10^{6}$ $c\bar{c}$ pairs.
The final D-meson pair yield is obtained by  multiplying by the fragmentation probability (larger for $D^0\bar{D}^0$ than for $D^+D^-$)  and the experimental D-meson reconstruction efficiency.
The latter can be improved by considering multiple decay modes, but the realistic estimate at the EIC is beyond the scope of this work.

\section{Conclusions}

In this paper, we have explored the possibility of performing quantum state tomography at the future EIC. We have argued that the spin density matrix $\rho$ of produced quark-antiquark pairs can be completely determined by measuring the azimuthal angular correlation of hadron and dihadron pairs generated by the Collins and dihadron fragmentation functions. The use of TMD  FFs not only aligns well with the mainstream research of the EIC, but also has practical advantages. If one tries to determine the spin density matrix elements through the production of heavy baryons such as $s\bar{s}\to \Lambda\bar{\Lambda}$ and $c\bar{c}\to \Lambda^+_c\bar{\Lambda}^-_c$ and their weak decays, one incurs a penalty factor of order  $10^{-4}\sim 10^{-6}$ due to the small fragmentation probabilities and small  branching ratios \cite{Kats:2023zxb,Fucilla:2026mkg}. This is a high price to pay for diffractive $q\bar{q}$ production whose cross section is typically in the picobarn range (see the last paragraph of Section VI), making  experimental measurements very challenging, especially for the charm quark. In contrast, the fragmentation probabilities  of favored channels such as $u\to \pi^+$ and $c\to D^+,D^0$ are order unity, and fortunately, the spin dependent fragmentation (for the transverse polarization) is not strongly suppressed compared to the spin independent fragmentation, see Fig.~\ref{fig:unintegrated-collins-2d}. Therefore, at least for diffractive production, the present approach appears to be the most practical way forward.

Our work is only the first step towards connecting the formal QCD calculations  \cite{Qi:2025onf,Fucilla:2025kit,Hatta:2025obw,Fucilla:2026mkg} with concrete experimental observables at the EIC. There are a number of directions for future research.   It is straightforward to extend the present calculation to diffractive production in the collinear factorization framework \cite{Hatta:2025obw}, as well as to inclusive $q\bar{q}$ production \cite{Qi:2025onf,Fucilla:2026mkg}. While the inclusive cross section is larger, the final states are more contaminated, and issues such as  decoherence \cite{Aoude:2025ovu,Gu:2025ijz,Agrawal:2026zwa,Cheng:2026zfb,Liu:2026ees} and the Sudakov resummation \cite{Catani:2014qha,Hatta:2021jcd} may be  severer. Since the longitudinal dihadron FF $G_1^\perp$ is unknown from the data,  insights from model calculations are useful \cite{Matevosyan:2017alv,Luo:2020wsg}.
At the moment, very little is known about higher twist effects in the fragmentation process beyond the quark mass effect (see the discussion around (\ref{34})). The MV model result points to the importance of  higher twist effects also in the initial state (i.e., the gluon saturation) if the target is a large nucleus.  Finally, in this paper we have worked in the symmetric (thrust axis) frame. The asymmetric frame avoids the complication of determining the thrust axis and may prove useful, or even necessary, for certain purposes.

\section*{Acknowledgments}

We thank Bowen Xiao for many discussions during the collaboration on  \cite{Fucilla:2026mkg} and in the subsequent development of this work. We also thank  Daniel Boer for correspondence.
Y.~G. and Y.~H. were supported by the U.S. Department
of Energy under Contract No. DE-SC0012704, and also by LDRD funds from Brookhaven Science Associates. The work of M.~F. is supported by the ULAM fellowship program of NAWA No. BNI/ULM/2024/1/00065 “Color glass condensate effective theory beyond the eikonal approximation”.

\appendix

\section{Spin density matrix elements from $\gamma$-matrices} \label{appA}

In this appendix, we give a detailed derivation of the formulas (\ref{cij}).
First recall the relation between four-dimensional spinors $u,v$ and the two-dimensional ones $\xi,\eta$
\beq
u_{\alpha}(p)=\begin{pmatrix} \sqrt{p\cdot \sigma}\xi_\alpha \\ \sqrt{p\cdot \bar{\sigma}}\xi_\alpha \end{pmatrix},
\qquad v_{\alpha'}(p')= \begin{pmatrix} \sqrt{p'\cdot \sigma}\tilde{\eta}_{-\alpha'} \\ -\sqrt{p'\cdot \bar{\sigma}}\tilde{\eta}_{-\alpha'} \end{pmatrix} = \begin{pmatrix} \sqrt{p\cdot \bar{\sigma}}\tilde{\eta}_{-\alpha'} \\ -\sqrt{p\cdot \sigma}\tilde{\eta}_{-\alpha'} \end{pmatrix}, \label{heli}
\eeq
Here, $\alpha$ and $\alpha'$ label the two spin states.  For the antiquark moving in the $-x^3$ direction, the `flipped spinors'
$\tilde{\eta}_{-\alpha'}=-i\sigma^2 (\eta_{\alpha'})^*$
are used \cite{Peskin:1995ev}. From (\ref{pp}), we can write
\beq
\sqrt{p\cdot \sigma}
= \sqrt{\frac{{\cal M}}{8}}\left( \sqrt{1+\beta}(\mathbb{I}-\sigma^z)+\sqrt{1-\beta}(\mathbb{I}+\sigma^z)\right)\equiv \sqrt{\frac{{\cal M}}{8}}K_+\cdot \sigma, \nn
\sqrt{p\cdot \bar{\sigma}} = \sqrt{\frac{{\cal M}}{8}}\left( \sqrt{1-\beta}(1-\sigma^z)+\sqrt{1+\beta}(1+\sigma^z)\right)\equiv \sqrt{\frac{{\cal M}}{8}}K_-\cdot \sigma.
\eeq
with $K_\pm^\mu=(\sqrt{1+\beta}+\sqrt{1-\beta},0,0,\pm(\sqrt{1+\beta}-\sqrt{1-\beta}))$.

Any 2 $\times$ 2 matrix can be expanded in the basis $\{ 1, \sigma^x, \sigma^y, \sigma^z \}$
\beq
\xi_\beta \xi^\dagger_{\alpha} &=& \frac{1}{2} \left( \xi^\dagger_\alpha \xi_\beta + (\xi^\dagger_\alpha \sigma^x \xi_\beta) \sigma^x + (\xi^\dagger_\alpha \sigma^y \xi_\beta) \sigma^y +(\xi^\dagger_\alpha \sigma^z \xi_\beta) \sigma^z \right) \; , \nn
\tilde{\eta}_{-\alpha'} \tilde{\eta}^\dagger_{-\beta'} &=& \frac{1}{2} \left( \tilde{\eta}^\dagger_{-\beta'} \tilde{\eta}_{-\alpha'} + (\tilde{\eta}^\dagger_{-\beta'} \sigma^x \tilde{\eta}_{-\alpha'}) \sigma^x + (\tilde{\eta}^\dagger_{-\beta'} \sigma^y \tilde{\eta}_{-\alpha'}) \sigma^y +(\tilde{\eta}^\dagger_{-\beta'} \sigma^z \tilde{\eta}_{-\alpha'}) \sigma^z \right) \nn
&=&\frac{1}{2} \left(  \eta_{\alpha'}^\dagger\eta_{\beta'} - (\eta^\dagger_{\alpha'} \sigma^x\eta_{\beta'}) \sigma^x - (\eta^\dagger_{\alpha'} \sigma^y\eta_{\beta'}) \sigma^y -(\eta^\dagger_{\alpha'} \sigma^z \eta_{\beta'}) \sigma^z \right)  \label{xixi}
\eeq
where we used the formula  $
\tilde{\eta}_{-\beta'}^\dagger \vec{\sigma}\tilde{\eta}_{-\alpha'} =
-\eta^\dagger_{\alpha'}\vec{\sigma}\eta_{\beta'}$.
We can then write in (\ref{trace}) ($a=x,y,z$)
\beq
u_\beta(k)\bar{u}_\alpha(k)= \frac{{\cal M}}{16}\left[\xi^\dagger_\alpha \xi_\beta \begin{pmatrix} K_+\cdot \sigma K_-\cdot \sigma & K_+ \cdot \sigma K_+\cdot \sigma \\ K_-\cdot \sigma K_-\cdot \sigma & K_-\cdot \sigma K_+\cdot \sigma \end{pmatrix}+\xi^\dagger_\alpha \sigma^a\xi_\beta \begin{pmatrix} K_+\cdot \sigma \sigma^a K_-\cdot \sigma & K_+ \cdot \sigma \sigma^a K_+\cdot \sigma \\ K_-\cdot \sigma \sigma^a K_-\cdot \sigma & K_-\cdot \sigma \sigma^a K_+\cdot \sigma \end{pmatrix}\right], \nonumber \\
\eeq

\beq
v_{\alpha'}(k')\bar{v}_{\beta'}(k')= \frac{{\cal M}}{16} \left[\eta^\dagger_{\alpha'} \eta_{\beta'}\begin{pmatrix} - K_-\cdot \sigma K_+\cdot \sigma & K_- \cdot \sigma K_-\cdot \sigma \\ K_+\cdot \sigma K_+\cdot \sigma & -K_+\cdot \sigma K_-\cdot \sigma \end{pmatrix}-\eta^\dagger_{\alpha'} \sigma^a\eta_{\beta'}\begin{pmatrix} -K_-\cdot \sigma \sigma^a K_+\cdot \sigma & K_- \cdot \sigma  \sigma^a K_-\cdot \sigma \\ K_+\cdot \sigma  \sigma^a K_+\cdot \sigma & -K_+\cdot \sigma  \sigma^a K_-\cdot \sigma \end{pmatrix}\right] . \nonumber \\
\eeq
Thus
\beq
A&=&\frac{{\cal M}^2}{64} {\rm Tr}\left[\bar{\Gamma} \begin{pmatrix} K_+\cdot \sigma K_-\cdot \sigma & K_+ \cdot \sigma K_+\cdot \sigma \\ K_-\cdot \sigma K_-\cdot \sigma & K_-\cdot \sigma K_+\cdot \sigma \end{pmatrix} \Gamma \begin{pmatrix} -K_-\cdot \sigma K_+\cdot \bar{\sigma} & K_- \cdot \sigma K_-\cdot \sigma \\ K_+\cdot \sigma K_+\cdot \sigma & -K_+\cdot \sigma K_-\cdot \sigma \end{pmatrix} \right] \nn
&=& {\rm Tr} \left[\bar{\Gamma} (\Slash k+m) \Gamma (\Slash k'-m)\right] \; ,
\eeq
\beq
AC^{ab}=-\frac{{\cal M}^2}{64}{\rm Tr}\Biggl[\bar{\Gamma}  \begin{pmatrix} K_+\cdot \sigma \sigma^a K_-\cdot \sigma & K_+ \cdot \sigma \sigma^a K_+\cdot \sigma \\ K_-\cdot \sigma \sigma^a K_-\cdot \sigma & K_-\cdot \sigma \sigma^a K_+\cdot \sigma \end{pmatrix} \Gamma\begin{pmatrix} -K_-\cdot \sigma \sigma^b K_+\cdot \sigma & K_- \cdot \sigma  \sigma^b K_-\cdot \sigma \\ K_+\cdot \sigma  \sigma^b K_+\cdot \sigma & -K_+\cdot \sigma  \sigma^b K_-\cdot \sigma \end{pmatrix} \Biggr] \; . \nonumber \\
\eeq
For $i,j=x,y$,
\beq
AC^{ij}&=&-\frac{{\cal M}^2}{64}{\rm Tr}\Biggl[\bar{\Gamma}  \begin{pmatrix} \sigma^i (K_-\cdot \sigma)^2  & \sigma^i K_- \cdot \sigma  K_+\cdot \sigma \\ \sigma^i K_+\cdot \sigma  K_-\cdot \sigma & \sigma^i (K_+\cdot \sigma)^2 \end{pmatrix} \Gamma\begin{pmatrix} -\sigma^j (K_+\cdot \sigma)^2 & \sigma^j K_+ \cdot \sigma  K_-\cdot \sigma \\ \sigma^j K_-\cdot \sigma  K_+\cdot \sigma & -\sigma^j (K_-\cdot \sigma)^2   \end{pmatrix} \Biggr] \nn
&=& -\frac{{\cal M}^2}{4}{\rm Tr}\Biggl[\bar{\Gamma}\begin{pmatrix} \sigma^i (1+\beta \sigma^z)  & \sqrt{1-\beta^2}\sigma^i  \\ \sqrt{1-\beta^2}\sigma^i  & \sigma^i (1-\beta\sigma^z) \end{pmatrix} \Gamma\begin{pmatrix} -\sigma^j (1-\beta\sigma^z) & \sqrt{1-\beta^2}\sigma^j  \\ \sqrt{1-\beta^2}\sigma^j & -\sigma^j (1+\beta \sigma^z) \end{pmatrix} \Biggr] \; .
\eeq
where we used $(K_\pm \cdot \sigma)^2=4(1\mp \beta \sigma^z)$ and $K_\pm\cdot \sigma K_\mp\cdot \sigma =4\sqrt{1-\beta^2}$.
Notice that
\beq
\begin{pmatrix} \sigma^i (1+\beta \sigma^z)  & \sqrt{1-\beta^2}\sigma^i  \\ \sqrt{1-\beta^2}\sigma^i  & \sigma^i (1-\beta\sigma^z) \end{pmatrix} =\gamma^i\gamma^5 \begin{pmatrix} \sqrt{1-\beta^2} & 1-\beta \sigma^z \\ 1+\beta\sigma^z & \sqrt{1-\beta^2} \end{pmatrix} &=&\gamma^i\gamma_5(\sqrt{1-\beta^2}+\gamma^0-\beta \gamma^z) \nn
&=& \frac{2}{{\cal M}}\gamma^i\gamma_5 (m+\Slash k) \; ,
\eeq
\beq
\begin{pmatrix} -\sigma^j (1-\beta\sigma^z) & \sqrt{1-\beta^2}\sigma^j  \\ \sqrt{1-\beta^2}\sigma^j & -\sigma^j (1+\beta \sigma^z) \end{pmatrix}=\gamma^j \gamma_5 \begin{pmatrix} \sqrt{1-\beta^2} & -(1+\beta\sigma^z) \\ -( 1-\beta\sigma^z) & \sqrt{1-\beta^2} \end{pmatrix} &=&\gamma^j\gamma_5 (\sqrt{1-\beta^2} -\gamma^0  -\beta\gamma^z) \nn
&=& \frac{2}{{\cal M}} \gamma^j\gamma_5 (m-\Slash k') \; .
\eeq
Similarly, if  $a=z$,
\beq
\begin{pmatrix} K_+\cdot \sigma \sigma^z K_-\cdot \sigma & K_+ \cdot \sigma \sigma^z K_+\cdot \sigma \\ K_-\cdot \sigma \sigma^z K_-\cdot \sigma & K_-\cdot \sigma \sigma^z K_+\cdot \sigma \end{pmatrix} =\frac{8}{{\cal M}}\gamma_5\gamma^0\gamma^3(\Slash k+m) \nn
\begin{pmatrix} -K_-\cdot \sigma \sigma^z K_+\cdot \sigma & K_- \cdot \sigma  \sigma^z K_-\cdot \sigma \\ K_+\cdot \sigma  \sigma^z K_+\cdot \sigma & -K_+\cdot \sigma  \sigma^z K_-\cdot \sigma \end{pmatrix} =\frac{8}{{\cal M}} \gamma_5\gamma^0\gamma^3(\Slash k'-m)  \label{mat}
\eeq

\section{Spin correlation matrix} \label{appB}

In this appendix, we reproduce the density matrix of a $q\bar{q}$ pair diffractively produced in $ep,eA$ scattering  \cite{Fucilla:2025kit}.
For the longitudinally polarized photon, the cross section is given by (see (\ref{omit}))
\begin{equation}
    A_L= \frac{8\alpha_{em}e_q^2\xi^2\bar{\xi}^2Q^2}{N_c} T_1^2,
\end{equation}
while
the $C$-matrix is the same as in the one-gluon exchange approximation  \cite{Qi:2025onf}
\beq
C^{yy}_L = 1, &\qquad&
C^{xx}_L=-C^{zz}_L = -\frac{k_\perp^2-(1-2\xi)^2m^2}{k_\perp^2+(1-2\xi)^2m^2}
= -\frac{{\cal M}^2\tan^2\theta-4m^2}{{\cal M}^2\tan^2\theta+4m^2},
\nn  && C^{xz}_L=C^{zx}_L =- \frac{2(1-2\xi)k_\perp m}{k_\perp^2+(1-2\xi)^2m^2}
=\frac{4{\cal M}m \tan\theta}{{\cal M}^2\tan^2\theta+4m^2}.
\label{cl}
\eeq
As observed in \cite{Qi:2025onf}, this density matrix represents a  maximally entangled state for any value of $k_\perp,\xi,m$.

For the transversely polarized photon, we have
\beq
A_T= \frac{2\alpha_{em}e_q^2}{N_c}((\xi^2+\bar{\xi}^2)k_\perp^2(T_1+T_2)^2 +m^2T_1^2)
\,, \label{at}
\eeq
where
\beq
T_1(k_\perp,\xi)=\int \frac{d^2q_\perp T(q_\perp)}{(k_\perp-q_\perp)^2+\xi\bar{\xi}Q^2+m^2}, \qquad T_2(k_\perp,\xi)= -\frac{1}{k_\perp^2}\int \frac{d^2q_\perp k_\perp\cdot q_\perp T(q_\perp)}{(k_\perp-q_\perp)^2+\xi\bar{\xi}Q^2+m^2}, \label{t1t2}
\eeq
and $T(q_\perp)$ is the so-called dipole T-matrix.  The $C$-matrix elements are given by
\beq
C^{xx}_T=2\xi\bar{\xi}k_\perp^2\frac{(k_\perp^4-m^4(1-2\xi)^2-2m^2k_\perp^2(\xi^2+\bar{\xi}^2))(T_1+T_2)^2 +4m^2k_\perp^2T_1(T_1+T_2) +2m^4T_1^2}{(k_\perp^2+m^2)(k_\perp^2+(1-2\xi)^2m^2)((\xi^2+\bar{\xi}^2)k_\perp^2(T_1+T_2)^2+m^2T_1^2)}, \label{cxx}
\eeq

\beq
C^{yy}_{T} = \frac{- 2\xi\bar{\xi}k_\perp^2(T_1+T_2)^2}{(\xi^2+\bar{\xi}^2)(T_1+T_2)^2k_\perp^2 +T_1^2m^2},
\eeq

\beq
C_T^{zz}
&=&\frac{1}{2((\xi^2+\bar{\xi}^2)(T_1+T_2)^2k_\perp^2+T_1^2m^2)}\Biggl[k_\perp^2(T_1+T_2)^2\left((1-2\xi)^2\frac{k_\perp^2-(1-2\xi)^2m^2}{k_\perp^2+(1-2\xi)^2m^2} +\frac{k_\perp^2-m^2}{k_\perp^2+m^2}\right) \\ && \qquad  + 4 m^2k_\perp^2T_1(T_1+T_2) \left(   \frac{(1-2\xi)^2}{k_\perp^2+(1-2\xi)^2m^2}  +\frac{1 }{k_\perp^2+m^2}  \right)  - \frac{2m^2(k_\perp^4-(1-2\xi)^2m^4)T_1^2}{(k_\perp^2+(1-2\xi)^2m^2)(k_\perp^2+m^2)}\Biggr] , \; \; \notag
\eeq
\beq
C_T^{xz}&=&
\frac{-k_\perp m}{(\xi^2+\bar{\xi}^2)(T_1+T_2)^2k_\perp^2+T_1^2m^2} \Biggl[k_\perp^2(T_1+T_2)^2\left(\frac{(1-2\xi)^3}{k_\perp^2+(1-2\xi)^2m^2} + \frac{1}{k_\perp^2+m^2}\right) \\
&& \quad   -T_1(T_1+T_2)\left((1-2\xi) \frac{k_\perp^2-(1-2\xi)^2m^2}{k_\perp^2+(1-2\xi)^2m^2} + \frac{k_\perp^2-m^2}{k_\perp^2+m^2}\right)- \frac{2(1-\xi)m^2(k_\perp^2+(1-2\xi)m^2)T_1^2}{(k_\perp^2+m^2)(k_\perp^2+(1-2\xi)^2m^2)}\Biggr], \notag \eeq
\beq
C_T^{zx}&=&
\frac{-k_\perp m}{(\xi^2+\bar{\xi}^2)(T_1+T_2)^2k_\perp^2+T_1^2m^2} \Biggl[k_\perp^2(T_1+T_2)^2\left(\frac{(1-2\xi)^3}{k_\perp^2+(1-2\xi)^2m^2} - \frac{1}{k_\perp^2+m^2}\right) \label{ckr}\\
&& \qquad  -T_1(T_1+T_2)\left((1-2\xi) \frac{k_\perp^2-(1-2\xi)^2m^2}{k_\perp^2+(1-2\xi)^2m^2} - \frac{k_\perp^2-m^2}{k_\perp^2+m^2}\right)+ \frac{2\xi m^2(k_\perp^2-(1-2\xi)m^2)T_1^2}{(k_\perp^2+m^2)(k_\perp^2+(1-2\xi)^2m^2)} \Biggr]. \notag
\eeq
Complicated as they may seem, these matrix elements exactly satisfy the relations (\ref{re1}) and (\ref{re2}).

To evaluate the above quantities, we employ two models for the dipole T-matrix: the Golec-Biernat-Wusthoff (GBW) model for the proton \cite{Golec-Biernat:1998zce}
and the McLerran-Venugopalan (MV) model for a heavy nucleus  \cite{McLerran:1993ni}. They are given by
\beq
T_{\rm GBW}(q_\perp)=\frac{N_c\sigma_0}{(2\pi)^4}\int d^2r_\perp e^{ir_\perp\cdot q_\perp}\left(1-e^{-\frac{r_\perp^2Q_s^2}{4}}\right)=\frac{N_c\sigma_0}{(2\pi)^2}\left(\delta^{(2)}(q_\perp) - \frac{1}{\pi Q_s^2}e^{-q_\perp^2/Q_s^2}\right), \label{gbw}
\eeq
with $Q_s=1$ GeV and  $\sigma_0=23$ mb, and
\beq
T_{\rm MV}(q_\perp)=\frac{A^{2/3}N_c\sigma_0}{(2\pi)^4}\int d^2r_\perp e^{ir_\perp\cdot q_\perp}\left(1-e^{-\frac{Q_{sA}^2r^2_\perp}{4}\ln \left(\frac{1}{r_\perp^2\Lambda^2}+e\right)}\right). \label{mv}
\eeq
with $\Lambda=0.24$ GeV.
We take  $Q_{sA}^2=1.5$ GeV$^2$, having a gold nucleus at the EIC in mind. Note that the prefactor drops out when computing $C^{ab}_{\rm DIS}$. Therefore, essentially the only difference between the two models is the logarithmic factor $\ln 1/r_\perp^2$ in the MV model.

The  maximally entangled state with $C_T^{yy}=-1$ found in \cite{Fucilla:2025kit}  is realized at $\xi=\bar{\xi}=\frac{1}{2}$ and $T_1=0$. It was argued in  \cite{Fucilla:2025kit} that $T_1$ crosses zero around the point (\ref{rem}) independently of models. However, we find that this is not the case for the MV model. The argument of \cite{Fucilla:2025kit} relies on the expansion of the integrand in  (\ref{t1t2}) ($\mu^2\equiv \frac{Q^2}{4}+m^2$)
\beq
T_1(k_\perp) \approx \frac{k_\perp^2-\mu^2}{(k_\perp^2+\mu^2)^3} \int d^2q_\perp q_\perp^2 T(q_\perp)+\cdots. \label{ex}
\eeq
We immediately recognize a zero at $k_\perp^2=\mu^2$, and we see from (\ref{cxx}) that $C^{xx}_T$ vanishes at this point (for $Q=0$).
However, the MV model features a perturbative tail $T_{\rm MV}(q_\perp)\sim 1/q_\perp^4$, so the integral is logarithmically divergent. Moreover, the higher order terms in (\ref{ex})  are even more divergent, meaning the breakdown of the expansion. Nevertheless, even in the MV model, $T_1(k_\perp)$ does cross zero at a numerically determined  point different from $k_\perp^2\approx  \mu^2$, and the maximally entangled state can be formed as shown in Fig.~\ref{fig:dis-charm-mv}. Incidentally, in collinear factorization, the second moment (\ref{ex}) is replaced by a collinear PDF (or GPD). Therefore, the zero at $k_\perp^2\approx \mu^2$ survives in this case, see \cite{Hatta:2025obw}.

\bibliography{ref}

@article{Hatta:2016dxp,
	author = "Hatta, Yoshitaka and Xiao, Bo-Wen and Yuan, Feng",
	title = "{Probing the Small- x Gluon Tomography in Correlated Hard Diffractive Dijet Production in Deep Inelastic Scattering}",
	eprint = "1601.01585",
	archivePrefix = "arXiv",
	primaryClass = "hep-ph",
	reportNumber = "YITP-16-1",
	doi = "10.1103/PhysRevLett.116.202301",
	journal = "Phys. Rev. Lett.",
	volume = "116",
	number = "20",
	pages = "202301",
	year = "2016"
}

@article{Salajegheh:2019nea,
	author = "Salajegheh, Maral and Moosavi Nejad, S. Mohammad and Soleymaninia, Maryam and Khanpour, Hamzeh and Atashbar Tehrani, S.",
	title = "{NNLO charmed-meson fragmentation functions and their uncertainties in the presence of meson mass corrections}",
	eprint = "1904.09832",
	archivePrefix = "arXiv",
	primaryClass = "hep-ph",
	doi = "10.1140/epjc/s10052-019-7521-x",
	journal = "Eur. Phys. J. C",
	volume = "79",
	number = "12",
	pages = "999",
	year = "2019"
}

@article{Anselmino:2015sxa,
	author = "Anselmino, M. and Boglione, M. and D'Alesio, U. and Gonzalez Hernandez, J. O. and Melis, S. and Murgia, F. and Prokudin, A.",
	title = "{Collins functions for pions from SIDIS and new $e^+e^-$ data: a first glance at their transverse momentum dependence}",
	eprint = "1510.05389",
	archivePrefix = "arXiv",
	primaryClass = "hep-ph",
	doi = "10.1103/PhysRevD.92.114023",
	journal = "Phys. Rev. D",
	volume = "92",
	number = "11",
	pages = "114023",
	year = "2015"
}

@article{Kang:2015msa,
	author = "Kang, Zhong-Bo and Prokudin, Alexei and Sun, Peng and Yuan, Feng",
	title = "{Extraction of Quark Transversity Distribution and Collins Fragmentation Functions with QCD Evolution}",
	eprint = "1505.05589",
	archivePrefix = "arXiv",
	primaryClass = "hep-ph",
	reportNumber = "JLAB-THY-15-2044",
	doi = "10.1103/PhysRevD.93.014009",
	journal = "Phys. Rev. D",
	volume = "93",
	number = "1",
	pages = "014009",
	year = "2016"
}

@article{Zeng:2023nnb,
	author = "Zeng, Chunhua and Dong, Hongxin and Liu, Tianbo and Sun, Peng and Zhao, Yuxiang",
	title = "{Role of sea quarks in the nucleon transverse spin}",
	eprint = "2310.15532",
	archivePrefix = "arXiv",
	primaryClass = "hep-ph",
	doi = "10.1103/PhysRevD.109.056002",
	journal = "Phys. Rev. D",
	volume = "109",
	number = "5",
	pages = "056002",
	year = "2024"
}

@article{Boussarie:2019ero,
	author = "Boussarie, R. and Grabovsky, A. V. and Szymanowski, L. and Wallon, S.",
	title = "{Towards a complete next-to-logarithmic description of forward exclusive diffractive dijet electroproduction at HERA: real corrections}",
	eprint = "1905.07371",
	archivePrefix = "arXiv",
	primaryClass = "hep-ph",
	reportNumber = "LPT-Orsay-19-22",
	doi = "10.1103/PhysRevD.100.074020",
	journal = "Phys. Rev. D",
	volume = "100",
	number = "7",
	pages = "074020",
	year = "2019"
}

@article{Pang:2026lsr,
	author = "Pang, Zhuoyi and Sznajder, Pawe{\l} and Szymanowski, Lech and Wagner, Jakub",
	title = "{Exclusive Quark and Gluon Dijet Production as Probes of GPDs at Collider Energies}",
	eprint = "2607.04482",
	archivePrefix = "arXiv",
	primaryClass = "hep-ph",
	month = "7",
	year = "2026"
}

@article{Cheng:2025cuv,
	author = "Cheng, Kun and Yan, Bin",
	title = "{Bell Inequality Violation of Light Quarks in Dihadron Pair Production at Lepton Colliders}",
	eprint = "2501.03321",
	archivePrefix = "arXiv",
	primaryClass = "hep-ph",
	reportNumber = "PITT-PACC-2414",
	doi = "10.1103/gmqz-v4cl",
	journal = "Phys. Rev. Lett.",
	volume = "135",
	number = "1",
	pages = "011902",
	year = "2025"
}

@article{White:2024nuc,
	author = "White, Chris D. and White, Martin J.",
	title = "{Magic states of top quarks}",
	eprint = "2406.07321",
	archivePrefix = "arXiv",
	primaryClass = "hep-ph",
	reportNumber = "ADP-24-10/T1249",
	doi = "10.1103/PhysRevD.110.116016",
	journal = "Phys. Rev. D",
	volume = "110",
	number = "11",
	pages = "116016",
	year = "2024"
}

@article{STAR:2025njp,
	author = "Aboona, B. E. and others",
	collaboration = "STAR",
	title = "{Measuring spin correlation between quarks during QCD confinement}",
	eprint = "2506.05499",
	archivePrefix = "arXiv",
	primaryClass = "hep-ex",
	doi = "10.1038/s41586-025-09920-0",
	journal = "Nature",
	volume = "650",
	number = "8100",
	pages = "65--71",
	year = "2026"
}

@article{PS185:2006yyx,
	author = "Paschke, K. D. and others",
	collaboration = "PS185",
	title = "{Experimental determination of the complete spin structure for anti-p p ---{\ensuremath{>}} anti-Lambda Lambda at p(anti-p) = 1.637-GeV/c}",
	eprint = "nucl-ex/0605025",
	archivePrefix = "arXiv",
	doi = "10.1103/PhysRevC.74.015206",
	journal = "Phys. Rev. C",
	volume = "74",
	pages = "015206",
	year = "2006"
}

@article{BESIII:2018cnd,
	author = "Ablikim, M. and others",
	collaboration = "BESIII",
	title = "{Polarization and Entanglement in Baryon-Antibaryon Pair Production in Electron-Positron Annihilation}",
	eprint = "1808.08917",
	archivePrefix = "arXiv",
	primaryClass = "hep-ex",
	doi = "10.1038/s41567-019-0494-8",
	journal = "Nature Phys.",
	volume = "15",
	pages = "631--634",
	year = "2019"
}

@article{Fabbrichesi:2021npl,
	author = "Fabbrichesi, M. and Floreanini, R. and Panizzo, G.",
	title = "{Testing Bell Inequalities at the LHC with Top-Quark Pairs}",
	eprint = "2102.11883",
	archivePrefix = "arXiv",
	primaryClass = "hep-ph",
	doi = "10.1103/PhysRevLett.127.161801",
	journal = "Phys. Rev. Lett.",
	volume = "127",
	number = "16",
	pages = "161801",
	year = "2021"
}

@article{Han:2023fci,
	author = "Han, Tao and Low, Matthew and Wu, Tong Arthur",
	title = "{Quantum entanglement and Bell inequality violation in semi-leptonic top decays}",
	eprint = "2310.17696",
	archivePrefix = "arXiv",
	primaryClass = "hep-ph",
	reportNumber = "PITT-PACC-2316",
	doi = "10.1007/JHEP07(2024)192",
	journal = "JHEP",
	volume = "07",
	pages = "192",
	year = "2024"
}

@article{Afik:2020onf,
	author = "Afik, Yoav and de Nova, Juan Ram{\'o}n Mu{\~n}oz",
	title = "{Entanglement and quantum tomography with top quarks at the LHC}",
	eprint = "2003.02280",
	archivePrefix = "arXiv",
	primaryClass = "quant-ph",
	doi = "10.1140/epjp/s13360-021-01902-1",
	journal = "Eur. Phys. J. Plus",
	volume = "136",
	number = "9",
	pages = "907",
	year = "2021"
}

@article{Dong:2023xiw,
	author = "Dong, Zhongtian and Gon{\c{c}}alves, Dorival and Kong, Kyoungchul and Navarro, Alberto",
	title = "{Entanglement and Bell inequalities with boosted tt{\textasciimacron}}",
	eprint = "2305.07075",
	archivePrefix = "arXiv",
	primaryClass = "hep-ph",
	doi = "10.1103/PhysRevD.109.115023",
	journal = "Phys. Rev. D",
	volume = "109",
	number = "11",
	pages = "115023",
	year = "2024"
}

@article{Gamberg:2022kdb,
	author = "Gamberg, Leonard and Malda, Michel and Miller, Joshua A. and Pitonyak, Daniel and Prokudin, Alexei and Sato, Nobuo",
	collaboration = "Jefferson Lab Angular Momentum (JAM)",
	title = "{Updated QCD global analysis of single transverse-spin asymmetries: Extracting H{\textasciitilde}, and the role of the Soffer bound and lattice QCD}",
	eprint = "2205.00999",
	archivePrefix = "arXiv",
	primaryClass = "hep-ph",
	reportNumber = "JLAB-THY-22-3604",
	doi = "10.1103/PhysRevD.106.034014",
	journal = "Phys. Rev. D",
	volume = "106",
	number = "3",
	pages = "034014",
	year = "2022"
}

@article{Boussarie:2023izj,
	author = "Boussarie, Renaud and others",
	title = "{TMD Handbook}",
	eprint = "2304.03302",
	archivePrefix = "arXiv",
	primaryClass = "hep-ph",
	reportNumber = "JLAB-THY-23-3780, LA-UR-21-20798, MIT-CTP/5386",
	month = "4",
	year = "2023"
}

@article{Horodecki:1995nsk,
	author = "Horodecki, R. and Horodecki, P. and Horodecki, M.",
	title = "{Violating Bell inequality by mixed

	spin-

	1
	2


	states: necessary and sufficient condition
	}",
	doi = "10.1016/0375-9601(95)00214-N",
	journal = "Phys. Lett. A",
	volume = "200",
	number = "5",
	pages = "340--344",
	year = "1995"
}

@article{Bernreuther:1993hq,
	author = "Bernreuther, Werner and Brandenburg, Arnd",
	title = "{Tracing CP violation in the production of top quark pairs by multiple TeV proton proton collisions}",
	eprint = "hep-ph/9312210",
	archivePrefix = "arXiv",
	reportNumber = "SLAC-PUB-6403, PITHA-93-43",
	doi = "10.1103/PhysRevD.49.4481",
	journal = "Phys. Rev. D",
	volume = "49",
	pages = "4481--4492",
	year = "1994"
}

@article{Chen:1994ar,
	author = "Chen, Kun and Goldstein, Gary R. and Jaffe, R. L. and Ji, Xiang-Dong",
	title = "{Probing quark fragmentation functions for spin 1/2 baryon production in unpolarized e+ e- annihilation}",
	eprint = "hep-ph/9410337",
	archivePrefix = "arXiv",
	reportNumber = "MIT-CTP-2365",
	doi = "10.1016/0550-3213(95)00193-V",
	journal = "Nucl. Phys. B",
	volume = "445",
	pages = "380--398",
	year = "1995"
}

@article{Artru:1995zu,
	author = "Artru, Xavier and Collins, John C.",
	title = "{Measuring transverse spin correlations by 4 particle correlations in e+ e- ---{\ensuremath{>}} 2 jets}",
	eprint = "hep-ph/9504220",
	archivePrefix = "arXiv",
	reportNumber = "PSU-TH-158, LYCEN-9511",
	doi = "10.1007/s002880050028",
	journal = "Z. Phys. C",
	volume = "69",
	pages = "277--286",
	year = "1996"
}

@article{Kneesch:2007ey,
	author = "Kneesch, T. and Kniehl, B. A. and Kramer, G. and Schienbein, I.",
	title = "{Charmed-meson fragmentation functions with finite-mass corrections}",
	eprint = "0712.0481",
	archivePrefix = "arXiv",
	primaryClass = "hep-ph",
	reportNumber = "DESY-07-215, LPSC-07-131",
	doi = "10.1016/j.nuclphysb.2008.02.015",
	journal = "Nucl. Phys. B",
	volume = "799",
	pages = "34--59",
	year = "2008"
}

@article{Cocuzza:2023vqs,
	author = "Cocuzza, C. and Metz, A. and Pitonyak, D. and Prokudin, A. and Sato, N. and Seidl, R.",
	collaboration = "Jefferson Lab Angular Momentum (JAM)",
	title = "{First simultaneous global QCD analysis of dihadron fragmentation functions and transversity parton distribution functions}",
	eprint = "2308.14857",
	archivePrefix = "arXiv",
	primaryClass = "hep-ph",
	reportNumber = "JLAB-THY-23-3901",
	doi = "10.1103/PhysRevD.109.034024",
	journal = "Phys. Rev. D",
	volume = "109",
	number = "3",
	pages = "034024",
	year = "2024"
}

@article{Cocuzza:2023oam,
	author = "Cocuzza, C. and Metz, A. and Pitonyak, D. and Prokudin, A. and Sato, N. and Seidl, R.",
	collaboration = "JAM",
	title = "{Transversity Distributions and Tensor Charges of the Nucleon: Extraction from Dihadron Production and Their Universal Nature}",
	eprint = "2306.12998",
	archivePrefix = "arXiv",
	primaryClass = "hep-ph",
	reportNumber = "JLAB-THY-23-3859",
	doi = "10.1103/PhysRevLett.132.091901",
	journal = "Phys. Rev. Lett.",
	volume = "132",
	number = "9",
	pages = "091901",
	year = "2024"
}

@article{deFlorian:2007aj,
	author = "de Florian, Daniel and Sassot, Rodolfo and Stratmann, Marco",
	title = "{Global analysis of fragmentation functions for pions and kaons and their uncertainties}",
	eprint = "hep-ph/0703242",
	archivePrefix = "arXiv",
	doi = "10.1103/PhysRevD.75.114010",
	journal = "Phys. Rev. D",
	volume = "75",
	pages = "114010",
	year = "2007"
}

@article{Belle:2008fdv,
	author = "Seidl, R. and others",
	collaboration = "Belle",
	title = "{Measurement of Azimuthal Asymmetries in Inclusive Production of Hadron Pairs in e+e- Annihilation at s**(1/2) = 10.58-GeV}",
	eprint = "0805.2975",
	archivePrefix = "arXiv",
	primaryClass = "hep-ex",
	reportNumber = "BELLE-2008-14, KEK-2008-8",
	doi = "10.1103/PhysRevD.78.032011",
	journal = "Phys. Rev. D",
	volume = "78",
	pages = "032011",
	year = "2008",
	note = "[Erratum: Phys.Rev.D 86, 039905 (2012)]"
}

@article{Golec-Biernat:1998zce,
	author = "Golec-Biernat, Krzysztof J. and Wusthoff, M.",
	title = "{Saturation effects in deep inelastic scattering at low Q**2 and its implications on diffraction}",
	eprint = "hep-ph/9807513",
	archivePrefix = "arXiv",
	reportNumber = "DTP-98-50",
	doi = "10.1103/PhysRevD.59.014017",
	journal = "Phys. Rev. D",
	volume = "59",
	pages = "014017",
	year = "1998"
}

@article{Aoude:2025ovu,
	author = "Aoude, Rafael and Barr, Alan J. and Maltoni, Fabio and Satrioni, Leonardo",
	title = "{Decoherence effects in entangled fermion pairs at colliders}",
	eprint = "2504.07030",
	archivePrefix = "arXiv",
	primaryClass = "quant-ph",
	doi = "10.1103/yjgh-f2gw",
	journal = "Phys. Rev. D",
	volume = "113",
	number = "7",
	pages = "076007",
	year = "2026"
}

@article{Cheng:2026zfb,
	author = "Cheng, Kun and Han, Tao and Singh, Harman and Su, Youle",
	title = "{Decoherence and More Coherence in the Radiative Decay of the $Z$ Boson}",
	eprint = "2607.12015",
	archivePrefix = "arXiv",
	primaryClass = "hep-ph",
	month = "7",
	year = "2026"
}

@article{Zhang:2026wvn,
	author = "Zhang, Hong-Wei and Cao, Xu and Feng, Tai-Fu",
	title = "{Controlling Quantum discord and steering in Electron-Positron Annihilation Using Polarized Beams}",
	eprint = "2605.19642",
	archivePrefix = "arXiv",
	primaryClass = "hep-ph",
	month = "5",
	year = "2026"
}

@article{Li:2026zum,
	author = "Li, Ying-Ying and Low, Ian and Wang, Yi-Lin and Yin, Zhewei",
	title = "{Quantum Magic in High Energy Collision}",
	eprint = "2608.19095",
	archivePrefix = "arXiv",
	primaryClass = "hep-ph",
	month = "8",
	year = "2026"
}

@article{Barr:2024djo,
	author = "Barr, Alan J. and Fabbrichesi, Marco and Floreanini, Roberto and Gabrielli, Emidio and Marzola, Luca",
	title = "{Quantum entanglement and Bell inequality violation at colliders}",
	eprint = "2402.07972",
	archivePrefix = "arXiv",
	primaryClass = "hep-ph",
	doi = "10.1016/j.ppnp.2024.104134",
	journal = "Prog. Part. Nucl. Phys.",
	volume = "139",
	pages = "104134",
	year = "2024"
}

@article{Qi:2025onf,
	author = "Qi, Wei and Guo, Zijing and Xiao, Bo-Wen",
	title = "{Studying maximal entanglement and Bell nonlocality at an electron-ion collider}",
	eprint = "2506.12889",
	archivePrefix = "arXiv",
	primaryClass = "hep-ph",
	doi = "10.1103/6ycn-x3yj",
	journal = "Phys. Rev. D",
	volume = "113",
	number = "5",
	pages = "054048",
	year = "2026"
}

@article{Afik:2025ejh,
	author = "Afik, Yoav and others",
	title = "{Quantum Information meets High-Energy Physics: Input to the update of the European Strategy for Particle Physics}",
	eprint = "2504.00086",
	archivePrefix = "arXiv",
	primaryClass = "hep-ph",
	month = "3",
	year = "2025"
}

@article{McLerran:1993ni,
	author = "McLerran, Larry D. and Venugopalan, Raju",
	title = "{Computing quark and gluon distribution functions for very large nuclei}",
	eprint = "hep-ph/9309289",
	archivePrefix = "arXiv",
	reportNumber = "TPI-MINN-93-44-T, NUC-MINN-93-24-T, HEP-UMN-TH-1220-93",
	doi = "10.1103/PhysRevD.49.2233",
	journal = "Phys. Rev. D",
	volume = "49",
	pages = "2233--2241",
	year = "1994"
}

@article{Fucilla:2026mkg,
	author = "Fucilla, Michael and Hatta, Yoshitaka and Xiao, Bo-Wen",
	title = "{Quantum entanglement in electron-nucleus collisions: Role of the linearly polarized gluon distribution}",
	eprint = "2604.11697",
	archivePrefix = "arXiv",
	primaryClass = "hep-ph",
	doi = "10.1103/6c93-2tgg",
	journal = "Phys. Rev. D",
	volume = "114",
	number = "3",
	pages = "034031",
	year = "2026"
}

@article{Matevosyan:2017liq,
	author = "Matevosyan, Hrayr H. and Kotzinian, Aram and Thomas, Anthony W.",
	title = "{Accessing Quark Helicity through Dihadron Studies}",
	eprint = "1712.06384",
	archivePrefix = "arXiv",
	primaryClass = "hep-ph",
	reportNumber = "ADP-17-42-T1048",
	doi = "10.1103/PhysRevLett.120.252001",
	journal = "Phys. Rev. Lett.",
	volume = "120",
	number = "25",
	pages = "252001",
	year = "2018"
}

@article{Matevosyan:2018icf,
	author = {Matevosyan, Hrayr H. and Bacchetta, Alessandro and Boer, Dani{\"e}l and Courtoy, Aurore and Kotzinian, Aram and Radici, Marco and Thomas, Anthony W.},
	title = "{Semi-inclusive production of two back-to-back hadron pairs in $e^+e^-$ annihilation revisited}",
	eprint = "1802.01578",
	archivePrefix = "arXiv",
	primaryClass = "hep-ph",
	reportNumber = "ADP-17-30-T1036, ADP-17-30/T1036",
	doi = "10.1103/PhysRevD.97.074019",
	journal = "Phys. Rev. D",
	volume = "97",
	number = "7",
	pages = "074019",
	year = "2018"
}

@article{CMS:2024pts,
	author = "Hayrapetyan, Aram and others",
	collaboration = "CMS",
	title = "{Observation of quantum entanglement in top quark pair production in proton{\textendash}proton collisions at $\sqrt{s} = 13$ TeV}",
	eprint = "2406.03976",
	archivePrefix = "arXiv",
	primaryClass = "hep-ex",
	reportNumber = "CMS-TOP-23-001, CERN-EP-2024-137",
	doi = "10.1088/1361-6633/ad7e4d",
	journal = "Rept. Prog. Phys.",
	volume = "87",
	number = "11",
	pages = "117801",
	year = "2024"
}

@article{Boer:2003ya,
	author = "Boer, Daniel and Jakob, Rainer and Radici, Marco",
	title = "{Interference fragmentation functions in electron positron annihilation}",
	eprint = "hep-ph/0302232",
	archivePrefix = "arXiv",
	doi = "10.1103/PhysRevD.67.094003",
	journal = "Phys. Rev. D",
	volume = "67",
	pages = "094003",
	year = "2003",
	note = "[Erratum: Phys.Rev.D 98, 039902 (2018)]"
}

@article{Bacchetta:2008wb,
	author = "Bacchetta, Alessandro and Ceccopieri, Federico Alberto and Mukherjee, Asmita and Radici, Marco",
	title = "{Asymmetries involving dihadron fragmentation functions: from DIS to e+e- annihilation}",
	eprint = "0812.0611",
	archivePrefix = "arXiv",
	primaryClass = "hep-ph",
	reportNumber = "JLAB-THY-08-917",
	doi = "10.1103/PhysRevD.79.034029",
	journal = "Phys. Rev. D",
	volume = "79",
	pages = "034029",
	year = "2009"
}

@article{Courtoy:2012ry,
	author = "Courtoy, Aurore and Bacchetta, Alessandro and Radici, Marco and Bianconi, Andrea",
	title = "{First extraction of Interference Fragmentation Functions from $e^+e^-$ data}",
	eprint = "1202.0323",
	archivePrefix = "arXiv",
	primaryClass = "hep-ph",
	doi = "10.1103/PhysRevD.85.114023",
	journal = "Phys. Rev. D",
	volume = "85",
	pages = "114023",
	year = "2012"
}

@article{Horodecki:1997vt,
	author = "Horodecki, Pawel",
	title = "{Separability criterion and inseparable mixed states with positive partial transposition}",
	eprint = "quant-ph/9703004",
	archivePrefix = "arXiv",
	doi = "10.1016/S0375-9601(97)00416-7",
	journal = "Phys. Lett. A",
	volume = "232",
	pages = "333",
	year = "1997"
}

@article{Peres:1996dw,
	author = "Peres, Asher",
	title = "{Separability criterion for density matrices}",
	eprint = "quant-ph/9604005",
	archivePrefix = "arXiv",
	doi = "10.1103/PhysRevLett.77.1413",
	journal = "Phys. Rev. Lett.",
	volume = "77",
	pages = "1413--1415",
	year = "1996"
}

@article{AbdulKhalek:2021gbh,
	author = "Abdul Khalek, R. and others",
	title = "{Science Requirements and Detector Concepts for the Electron-Ion Collider}: {EIC Yellow Report}",
	eprint = "2103.05419",
	archivePrefix = "arXiv",
	primaryClass = "physics.ins-det",
	reportNumber = "BNL-220990-2021-FORE, JLAB-PHY-21-3198, LA-UR-21-20953",
	doi = "10.1016/j.nuclphysa.2022.122447",
	journal = "Nucl. Phys. A",
	volume = "1026",
	pages = "122447",
	year = "2022"
}

@article{ATLAS:2023fsd,
	author = "Aad, Georges and others",
	collaboration = "ATLAS",
	title = "{Observation of quantum entanglement with top quarks at the ATLAS detector}",
	eprint = "2311.07288",
	archivePrefix = "arXiv",
	primaryClass = "hep-ex",
	reportNumber = "CERN-EP-2023-230",
	doi = "10.1038/s41586-024-07824-z",
	journal = "Nature",
	volume = "633",
	number = "8030",
	pages = "542--547",
	year = "2024"
}

@article{BaBar:2013jdt,
	author = "Lees, J. P. and others",
	collaboration = "BaBar",
	title = "{Measurement of Collins asymmetries in inclusive production of charged pion pairs in $e^+e^-$ annihilation at BABAR}",
	eprint = "1309.5278",
	archivePrefix = "arXiv",
	primaryClass = "hep-ex",
	reportNumber = "BABAR-PUB-13-012, SLAC-PUB-15700",
	doi = "10.1103/PhysRevD.90.052003",
	journal = "Phys. Rev. D",
	volume = "90",
	number = "5",
	pages = "052003",
	year = "2014"
}

@article{Hatta:2025obw,
	author = "Hatta, Yoshitaka and Schoenleber, Jakob",
	title = "{Probing quantum entanglement with generalized parton distributions at the Electron-Ion Collider}",
	eprint = "2511.04537",
	archivePrefix = "arXiv",
	primaryClass = "hep-ph",
	doi = "10.1103/qdb2-k2nh",
	journal = "Phys. Rev. D",
	volume = "113",
	number = "9",
	pages = "094016",
	year = "2026"
}

@article{Bloss:2026yrf,
	author = "Bloss, Henry and Hobbs, T. J. and McGinnis, Navin",
	title = "{Hadron Structure from the Hierarchy of Quantum Correlations in Deep-Inelastic Scattering}",
	eprint = "2607.28724",
	archivePrefix = "arXiv",
	primaryClass = "hep-ph",
	reportNumber = "ANL-205269",
	month = "7",
	year = "2026"
}

@article{H1:2020lzc,
	author = "Andreev, V. and others",
	collaboration = "H1",
	title = "{Measurement of Exclusive $\pi^{+}\pi^{-}$ and $\rho^0$ Meson Photoproduction at HERA}",
	eprint = "2005.14471",
	archivePrefix = "arXiv",
	primaryClass = "hep-ex",
	reportNumber = "DESY-20-080",
	doi = "10.1140/epjc/s10052-020-08587-3",
	journal = "Eur. Phys. J. C",
	volume = "80",
	number = "12",
	pages = "1189",
	year = "2020"
}

@article{Bianconi:1999cd,
	author = "Bianconi, A. and Boffi, S. and Jakob, R. and Radici, M.",
	title = "{Two hadron interference fragmentation functions. Part 1. General framework}",
	eprint = "hep-ph/9907475",
	archivePrefix = "arXiv",
	doi = "10.1103/PhysRevD.62.034008",
	journal = "Phys. Rev. D",
	volume = "62",
	pages = "034008",
	year = "2000"
}

@article{CMS:2024zkc,
	author = "Hayrapetyan, Aram and others",
	collaboration = "CMS",
	title = "{Measurements of polarization and spin correlation and observation of entanglement in top quark pairs using lepton+jets events from proton-proton collisions at s=13{\,}{\,}TeV}",
	eprint = "2409.11067",
	archivePrefix = "arXiv",
	primaryClass = "hep-ex",
	reportNumber = "CMS-TOP-23-007, CERN-EP-2024-231",
	doi = "10.1103/PhysRevD.110.112016",
	journal = "Phys. Rev. D",
	volume = "110",
	number = "11",
	pages = "112016",
	year = "2024"
}

@article{Galanti:2015pqa,
	author = "Galanti, Mario and Giammanco, Andrea and Grossman, Yuval and Kats, Yevgeny and Stamou, Emmanuel and Zupan, Jure",
	title = "{Heavy baryons as polarimeters at colliders}",
	eprint = "1505.02771",
	archivePrefix = "arXiv",
	primaryClass = "hep-ph",
	reportNumber = "CP3-15-12",
	doi = "10.1007/JHEP11(2015)067",
	journal = "JHEP",
	volume = "11",
	pages = "067",
	year = "2015"
}

@book{Peskin:1995ev,
	author = "Peskin, Michael E. and Schroeder, Daniel V.",
	title = "{An Introduction to quantum field theory}",
	doi = "10.1201/9780429503559",
	isbn = "978-0-201-50397-5, 978-0-429-50355-9, 978-0-429-49417-8",
	publisher = "Addison-Wesley",
	address = "Reading, USA",
	year = "1995"
}

@article{Metz:2016swz,
	author = "Metz, Andreas and Vossen, Anselm",
	title = "{Parton Fragmentation Functions}",
	eprint = "1607.02521",
	archivePrefix = "arXiv",
	primaryClass = "hep-ex",
	doi = "10.1016/j.ppnp.2016.08.003",
	journal = "Prog. Part. Nucl. Phys.",
	volume = "91",
	pages = "136--202",
	year = "2016"
}

@article{Xiao:2026tbs,
	author = "Xiao, Bo-Wen and Zhao, Yuxiang and Zhou, Jian",
	title = "{Physics of the Electron{\textendash}Ion Collider in China}",
	eprint = "2608.11712",
	archivePrefix = "arXiv",
	primaryClass = "hep-ph",
	doi = "10.1016/j.ppnp.2026.104264",
	journal = "Prog. Part. Nucl. Phys.",
	volume = "151",
	pages = "104264",
	year = "2026"
}

@article{Braun:2005rg,
	author = "Braun, V. M. and Ivanov, D. Yu.",
	title = "{Exclusive diffractive electroproduction of dijets in collinear factorization}",
	eprint = "hep-ph/0505263",
	archivePrefix = "arXiv",
	doi = "10.1103/PhysRevD.72.034016",
	journal = "Phys. Rev. D",
	volume = "72",
	pages = "034016",
	year = "2005"
}

@article{Bartels:1996ne,
	author = {Bartels, Jochen and Lotter, H. and W{\"u}sthoff, M.},
	title = "{Quark-antiquark production in DIS diffractive dissociation}",
	eprint = "hep-ph/9602363",
	archivePrefix = "arXiv",
	reportNumber = "DESY-96-026, ANL-HEP-PR-96-11",
	doi = "10.1016/0370-2693(96)00412-1",
	journal = "Phys. Lett. B",
	volume = "379",
	pages = "239--248",
	year = "1996",
	note = "[Erratum: Phys.Lett.B 382, 449--449 (1996)]"
}

@article{Nikolaev:2003zf,
	author = "Nikolaev, N. N. and Schafer, W. and Zakharov, B. G. and Zoller, V. R.",
	title = "{Nonlinear k-perpendicular factorization for forward dijets in DIS off nuclei in the saturation regime}",
	eprint = "hep-ph/0303024",
	archivePrefix = "arXiv",
	doi = "10.1134/1.1618332",
	journal = "J. Exp. Theor. Phys.",
	volume = "97",
	pages = "441--465",
	year = "2003"
}

@article{Bernal:2023jba,
	author = "Bernal, Alexander",
	title = "{Quantum tomography of helicity states for general scattering processes}",
	eprint = "2310.10838",
	archivePrefix = "arXiv",
	primaryClass = "hep-ph",
	doi = "10.1103/PhysRevD.109.116007",
	journal = "Phys. Rev. D",
	volume = "109",
	number = "11",
	pages = "116007",
	year = "2024"
}

@article{Altinoluk:2015dpi,
	author = "Altinoluk, Tolga and Armesto, N{\'e}stor and Beuf, Guillaume and Rezaeian, Amir H.",
	title = "{Diffractive Dijet Production in Deep Inelastic Scattering and Photon-Hadron Collisions in the Color Glass Condensate}",
	eprint = "1511.07452",
	archivePrefix = "arXiv",
	primaryClass = "hep-ph",
	doi = "10.1016/j.physletb.2016.05.032",
	journal = "Phys. Lett. B",
	volume = "758",
	pages = "373--383",
	year = "2016"
}

@article{Liu:2026ees,
	author = "Liu, Feng and Tu, Zhoudunming",
	title = "{Quantum decoherence of hyperon spin correlations in QCD hadronization}",
	eprint = "2606.17240",
	archivePrefix = "arXiv",
	primaryClass = "hep-ph",
	month = "6",
	year = "2026"
}

@article{Diehl:1994wz,
	author = "Diehl, M.",
	title = "{Diffractive production of dijets at HERA}",
	eprint = "hep-ph/9407399",
	archivePrefix = "arXiv",
	reportNumber = "DAMTP-94-60",
	doi = "10.1007/BF01496592",
	journal = "Z. Phys. C",
	volume = "66",
	pages = "181--194",
	year = "1995"
}

@article{Boer:1997mf,
	author = "Boer, Daniel and Jakob, R. and Mulders, P. J.",
	title = "{Asymmetries in polarized hadron production in e+ e- annihilation up to order 1/Q}",
	eprint = "hep-ph/9702281",
	archivePrefix = "arXiv",
	reportNumber = "NIKHEF-97-008, VUTH-97-3",
	doi = "10.1016/S0550-3213(97)00456-2",
	journal = "Nucl. Phys. B",
	volume = "504",
	pages = "345--380",
	year = "1997"
}

@article{Anselmino:2007fs,
	author = "Anselmino, M. and Boglione, M. and D'Alesio, U. and Kotzinian, A. and Murgia, F. and Prokudin, A. and Turk, C.",
	title = "{Transversity and Collins functions from SIDIS and e+ e- data}",
	eprint = "hep-ph/0701006",
	archivePrefix = "arXiv",
	doi = "10.1103/PhysRevD.75.054032",
	journal = "Phys. Rev. D",
	volume = "75",
	pages = "054032",
	year = "2007"
}

@article{Pitonyak:2013dsu,
	author = "Pitonyak, D. and Schlegel, M. and Metz, A.",
	title = "{Polarized hadron pair production from electron-positron annihilation}",
	eprint = "1310.6240",
	archivePrefix = "arXiv",
	primaryClass = "hep-ph",
	reportNumber = "RBRC-1046",
	doi = "10.1103/PhysRevD.89.054032",
	journal = "Phys. Rev. D",
	volume = "89",
	number = "5",
	pages = "054032",
	year = "2014"
}

@article{Severi:2021cnj,
	author = "Severi, Claudio and Boschi, Cristian Degli Esposti and Maltoni, Fabio and Sioli, Maximiliano",
	title = "{Quantum tops at the LHC: from entanglement to Bell inequalities}",
	eprint = "2110.10112",
	archivePrefix = "arXiv",
	primaryClass = "hep-ph",
	doi = "10.1140/epjc/s10052-022-10245-9",
	journal = "Eur. Phys. J. C",
	volume = "82",
	number = "4",
	pages = "285",
	year = "2022"
}

@article{Hayward:2021psm,
	author = "Hayward, T. B. and others",
	title = "{Observation of Beam Spin Asymmetries in the Process $ep\rightarrow{e}^{'}{\pi}^{+}{\pi}^{-}X$ with CLAS12}",
	eprint = "2101.04842",
	archivePrefix = "arXiv",
	primaryClass = "hep-ex",
	reportNumber = "JLAB-PHY-21-3307",
	doi = "10.1103/PhysRevLett.126.152501",
	journal = "Phys. Rev. Lett.",
	volume = "126",
	pages = "152501",
	year = "2021"
}

@article{Luo:2020wsg,
	author = "Luo, Xuan and Sun, Hao and Xie, Yi-Ling",
	title = "{Single spin asymmetry $A_{UL}^{\sin(\phi_h-\phi_R)}$ in dihadron semi-inclusive DIS}",
	eprint = "2003.03770",
	archivePrefix = "arXiv",
	primaryClass = "hep-ph",
	doi = "10.1103/PhysRevD.101.054020",
	journal = "Phys. Rev. D",
	volume = "101",
	number = "5",
	pages = "054020",
	year = "2020"
}

@article{Matevosyan:2017alv,
	author = "Matevosyan, Hrayr H. and Kotzinian, Aram and Thomas, Anthony W.",
	title = "{Dihadron fragmentation functions in the quark-jet model: Longitudinally polarized quarks}",
	eprint = "1707.04999",
	archivePrefix = "arXiv",
	primaryClass = "hep-ph",
	reportNumber = "ADP-17-30-T1036",
	doi = "10.1103/PhysRevD.96.074010",
	journal = "Phys. Rev. D",
	volume = "96",
	number = "7",
	pages = "074010",
	year = "2017"
}

@article{Collins:1992kk,
	author = "Collins, John C.",
	title = "{Fragmentation of transversely polarized quarks probed in transverse momentum distributions}",
	eprint = "hep-ph/9208213",
	archivePrefix = "arXiv",
	reportNumber = "PSU-TH-102",
	doi = "10.1016/0550-3213(93)90262-N",
	journal = "Nucl. Phys. B",
	volume = "396",
	pages = "161--182",
	year = "1993"
}

@article{Boer:2008fr,
	author = "Boer, Daniel",
	title = "{Angular dependences in inclusive two-hadron production at BELLE}",
	eprint = "0804.2408",
	archivePrefix = "arXiv",
	primaryClass = "hep-ph",
	doi = "10.1016/j.nuclphysb.2008.06.011",
	journal = "Nucl. Phys. B",
	volume = "806",
	pages = "23--67",
	year = "2009"
}

@article{Kanachova:2026ywy,
	author = "Kanachova, Veronica and Liu, Feng and Tu, Zhoudunming",
	title = "{Qubit-Qutrit Quantum Tomography of hadronic $Λϕ$ and $ΛK^{\ast 0}$ systems}",
	eprint = "2609.11151",
	archivePrefix = "arXiv",
	primaryClass = "hep-ph",
	month = "9",
	year = "2026"
}

@article{Lin:2025eci,
	author = "Lin, Shi-Jia and Liu, Ming-Jun and Shao, Ding Yu and Wei, Shu-Yi",
	title = "{Spin correlations and Bell nonlocality in $ \Lambda \overline{\Lambda} $ pair production from e$^{+}$e$^{−}$ collisions with a thrust cut}",
	eprint = "2507.15387",
	archivePrefix = "arXiv",
	primaryClass = "hep-ph",
	doi = "10.1007/JHEP11(2025)082",
	journal = "JHEP",
	volume = "11",
	pages = "082",
	year = "2025"
}

@article{BESIII:2015fyw,
	author = "Ablikim, M. and others",
	collaboration = "BESIII",
	title = "{Measurement of azimuthal asymmetries in inclusive charged dipion production in $e^+e^-$ annihilations at $\sqrt{s}$ = 3.65 GeV}",
	eprint = "1507.06824",
	archivePrefix = "arXiv",
	primaryClass = "hep-ex",
	doi = "10.1103/PhysRevLett.116.042001",
	journal = "Phys. Rev. Lett.",
	volume = "116",
	number = "4",
	pages = "042001",
	year = "2016"
}

@article{Cao:2026zqo,
	author = "Cao, Haotian and Guo, Yuxun and Hatta, Yoshitaka and Schoenleber, Jakob",
	title = "{Three-qubit entanglement in the Bethe-Heitler process}",
	eprint = "2608.18030",
	archivePrefix = "arXiv",
	primaryClass = "quant-ph",
	month = "8",
	year = "2026"
}

@article{Kats:2023zxb,
	author = "Kats, Yevgeny and Uzan, David",
	title = "{Prospects for measuring quark polarization and spin correlations in $b\overline{b }$ and $c\overline{c }$ samples at the LHC}",
	eprint = "2311.08226",
	archivePrefix = "arXiv",
	primaryClass = "hep-ph",
	doi = "10.1007/JHEP03(2024)063",
	journal = "JHEP",
	volume = "03",
	pages = "063",
	year = "2024"
}

@article{Mantysaari:2019csc,
	author = {M{\"a}ntysaari, Heikki and Mueller, Niklas and Schenke, Bj{\"o}rn},
	title = "{Diffractive Dijet Production and Wigner Distributions from the Color Glass Condensate}",
	eprint = "1902.05087",
	archivePrefix = "arXiv",
	primaryClass = "hep-ph",
	doi = "10.1103/PhysRevD.99.074004",
	journal = "Phys. Rev. D",
	volume = "99",
	number = "7",
	pages = "074004",
	year = "2019"
}

@article{Gu:2025ijz,
	author = "Gu, Jiayin and Lin, Shi-Jia and Shao, Ding Yu and Wang, Lian-Tao and Yang, Si-Xiang",
	title = "{Decoherence in high energy collisions as renormalization group flow}",
	eprint = "2510.13951",
	archivePrefix = "arXiv",
	primaryClass = "hep-ph",
	month = "10",
	year = "2025"
}

@article{Agrawal:2026zwa,
	author = "Agrawal, Sanskriti and Zahoor, Muneeb and Abir, Raktim",
	title = "{Soft-Radiation-Induced Decoherence of Heavy-Quark Spin Entanglement at the Electron-Ion Collider}",
	eprint = "2606.29944",
	archivePrefix = "arXiv",
	primaryClass = "hep-ph",
	month = "6",
	year = "2026"
}

@article{Cheng:2025zaw,
	author = "Cheng, Kun and Han, Tao and Trifinopoulos, Sokratis",
	title = "{Quantum information at the electron-ion collider}",
	eprint = "2510.23773",
	archivePrefix = "arXiv",
	primaryClass = "hep-ph",
	reportNumber = "PITT-PACC-2509, CERN-TH-2025-205, MIT-CTP/5943",
	doi = "10.1103/xytf-nhkh",
	journal = "Phys. Rev. D",
	volume = "114",
	number = "5",
	pages = "L051301",
	year = "2026"
}

@article{Catani:2014qha,
	author = "Catani, Stefano and Grazzini, Massimiliano and Torre, A.",
	title = "{Transverse-momentum resummation for heavy-quark hadroproduction}",
	eprint = "1408.4564",
	archivePrefix = "arXiv",
	primaryClass = "hep-ph",
	reportNumber = "ZU-TH-28-14",
	doi = "10.1016/j.nuclphysb.2014.11.019",
	journal = "Nucl. Phys. B",
	volume = "890",
	pages = "518--538",
	year = "2014"
}

@article{Hatta:2021jcd,
	author = "Hatta, Yoshitaka and Xiao, Bo-Wen and Yuan, Feng and Zhou, Jian",
	title = "{Azimuthal angular asymmetry of soft gluon radiation in jet production}",
	eprint = "2106.05307",
	archivePrefix = "arXiv",
	primaryClass = "hep-ph",
	doi = "10.1103/PhysRevD.104.054037",
	journal = "Phys. Rev. D",
	volume = "104",
	number = "5",
	pages = "054037",
	year = "2021"
}

@article{Hatta:2026dqs,
	author = "Hatta, Yoshitaka and Mart{\'\i}nez-Fern{\'a}ndez, V{\'\i}ctor",
	title = "{Photon-nucleon entanglement in Compton scattering at low and high energies}",
	eprint = "2608.05330",
	archivePrefix = "arXiv",
	primaryClass = "hep-ph",
	month = "8",
	year = "2026"
}

@article{Fucilla:2025kit,
	author = "Fucilla, Michael and Hatta, Yoshitaka",
	title = "{Spin-spin entanglement in diffractive heavy-quark production}",
	eprint = "2509.05267",
	archivePrefix = "arXiv",
	primaryClass = "hep-ph",
	doi = "10.1103/gbk8-z3dd",
	journal = "Phys. Rev. D",
	volume = "113",
	number = "3",
	pages = "L031504",
	year = "2026"
}

\end{document}